\documentclass[UKenglish,cleveref, autoref, thm-restate]{article}
\usepackage{blindtext}
\usepackage[a4paper, total={6in, 10in}]{geometry}
\usepackage{hyperref}
\usepackage{cite}
\usepackage{amsmath,amssymb,amsfonts}
\usepackage{subcaption}
\usepackage{textcomp}
\usepackage{xcolor}
\usepackage{glossaries}
\usepackage{svg}
\usepackage{algorithm}
\usepackage{multirow}
\newacronym{pbs}{PBS}{Proposer-Builder Separation}
\newacronym{ofc}{OFC}{Order Flow Composition}
\newacronym{bft}{BFT}{Byzantine Fault Tolerant}
\newacronym{pos}{PoS}{Proof-of-Stake}
\newacronym{pow}{PoW}{Proof-of-Work}
\newacronym{mev}{MEV}{Maximal Extractable Value}
\newacronym[shortplural=DEXes]{dex}{DEX}{Decentralized Exchange}
\newacronym[shortplural=CEXes]{cex}{CEX}{Centralized Exchange} 
\newacronym{amm}{AMM}{Automated Market Maker}
\newacronym{defi}{DeFi}{Decentralized Finance}
\newacronym{pga}{PGA}{Priority Gas Auction}
\newacronym{tob}{ToB}{Top-of-Block}
\newacronym{bob}{BoB}{Body-of-Block}
\newacronym{eob}{EoB}{End-of-Block}
\newacronym{ofa}{OFA}{Order Flow Auction}
\newacronym{0xce91228789b57deb45e66ca10ff648385fe7093b}{0xCe91228789B57DEb45e66Ca10Ff648385fE7093b}{MEV Blocker Rebates Safe}
\newacronym{rfq}{RFQ}{Request For Quote}
\newacronym{pm}{PM}{Profit Margin}
\newacronym{eoa}{EOA}{Externally Owned Account}
\newacronym{eof}{EOF}{Exclusive Order Flow}
\newacronym{msof}{MSOF}{Most Significant Order Flow}
\newacronym{jit}{JIT}{Just-In-Time}
\newacronym{lda}{LDA}{Linear Discriminant Analysis}
\newacronym{da}{DA}{Decoding Accuracy}
\newacronym{rpc}{RPC}{Remote Procedure Call}
\newacronym{brt}{BRT}{\texttt{beaverbuild}, \texttt{rysnc}, and \texttt{Titan}}
\newacronym{ep}{EP}{Exclusive Provider}
\newacronym{tee}{TEE}{Trusted Execution Environment}
\newacronym{il}{IL}{Inclusion List}
\newacronym{aps}{APS}{Attestor-Proposer Separation}
\newacronym{ea}{EA}{Execution Auction}
\newacronym{cr}{CR}{Censorship-Resistance}
\newacronym{hft}{HFT}{High-Frequency Trading}
\usepackage{graphicx}
\usepackage{booktabs,arydshln}
\usepackage{pdflscape}
\usepackage{subcaption}
\usepackage{ragged2e}
\usepackage{cleveref}
\usepackage{siunitx}
\usepackage{authblk}

\makeatletter
\def\adl@drawiv#1#2#3{%
        \hskip.5\tabcolsep
        \xleaders#3{#2.5\@tempdimb #1{1}#2.5\@tempdimb}%
                #2\z@ plus1fil minus1fil\relax
        \hskip.5\tabcolsep}
\newcommand{\cdashlinelr}[1]{%
  \noalign{\vskip\aboverulesep
           \global\let\@dashdrawstore\adl@draw
           \global\let\adl@draw\adl@drawiv}
  \cdashline{#1}
  \noalign{\global\let\adl@draw\@dashdrawstore
           \vskip\belowrulesep}}
\makeatother

\title{Who Wins Ethereum Block Building Auctions and Why?}

\author[1]{Burak Öz}
\author[2]{Danning Sui}
\author[3]{Thomas Thiery}
\author[1]{Florian Matthes}
\affil[1]{Technical University of Munich, Garching, Germany}
\affil[2]{Flashbots, San Francisco, CA, USA}
\affil[3]{Ethereum Foundation, Lisbon, Portugal}

\date{}

\begin{document}

\maketitle

\begin{abstract}
The MEV-Boost block auction contributes approximately \SI{90}{\%} of all Ethereum blocks. Between October 2023 and March 2024, only three builders produced \SI{80}{\%} of them, highlighting the concentration of power within the block builder market. To foster competition and preserve Ethereum's decentralized ethos and censorship-resistance properties, understanding the dominant players' competitive edges is essential.

In this paper, we identify features that play a significant role in builders' ability to win blocks and earn profits by conducting a comprehensive empirical analysis of MEV-Boost auctions over a six-month period. We reveal that block market share positively correlates with order flow diversity, while profitability correlates with access to order flow from Exclusive Providers, such as integrated searchers and external providers with exclusivity deals. Additionally, we show a positive correlation between market share and profit margin among the top ten builders, with features such as exclusive signal, non-atomic arbitrages, and Telegram bot flow strongly correlating with both metrics. This highlights a ``chicken-and-egg'' problem where builders need differentiated order flow to profit, but only receive such flow if they have a significant market share. Overall, this work provides an in-depth analysis of the key features driving the builder market towards centralization and offers valuable insights for designing further iterations of Ethereum block auctions, preserving Ethereum's censorship resistance properties.
\end{abstract}

\section{Introduction}
Ethereum blockchain grows by one block every 12 seconds if the block proposal slot is not missed. The consensus specifications \cite{noauthor_ethereumconsensus-specs_2024} expect the selected consensus participant, the \textit{validator}, to build a block and propose it to the rest of the network. However, nearly \SI{90}{\%} \cite{wahrstatter_mev-boost_nodate} of all new blocks are produced via the MEV-Boost\cite{noauthor_flashbotsmev-boost_2024} block building auction, which is a \gls{pbs} \cite{vitalik_pbs} implementation. In this mechanism, the proposer acts as an auctioneer, selling their block building rights, while a set of out-of-protocol entities, referred to as \textit{builders}, compete by sending bids to intermediaries known as \textit{relays}. The proposed MEV-Boost block is signed by the respective proposer, while the payload is constructed by the winning builder. By separating block proposing and building tasks, MEV-Boost aims to avoid validator centralization caused by the sophistication of running complex strategies to extract \gls{mev} \cite{daian_flash_2020}.

As an out-of-protocol instantiation of \gls{pbs}, MEV-Boost preserves decentralization on the consensus layer as validators have uniform access to \gls{mev} rewards \cite{bahrani_centralization_2024}. However, it concentrates power within the builder market due to economies of scale. Between October 2023 and March 2024, only three builders, \texttt{beaverbuild}, \texttt{rsync}, and \texttt{Titan}, produced approximately \SI{80}{\%} of all MEV-Boost blocks. With a small set of builders creating Ethereum blocks, the network becomes vulnerable to censorship \cite{fox_censorship_2023, wahrstatter_blockchain_2023, wahrstatter_censorshippics_nodate}.

\begin{figure}[t!]
    \centering
    \includegraphics[width=0.75\linewidth]{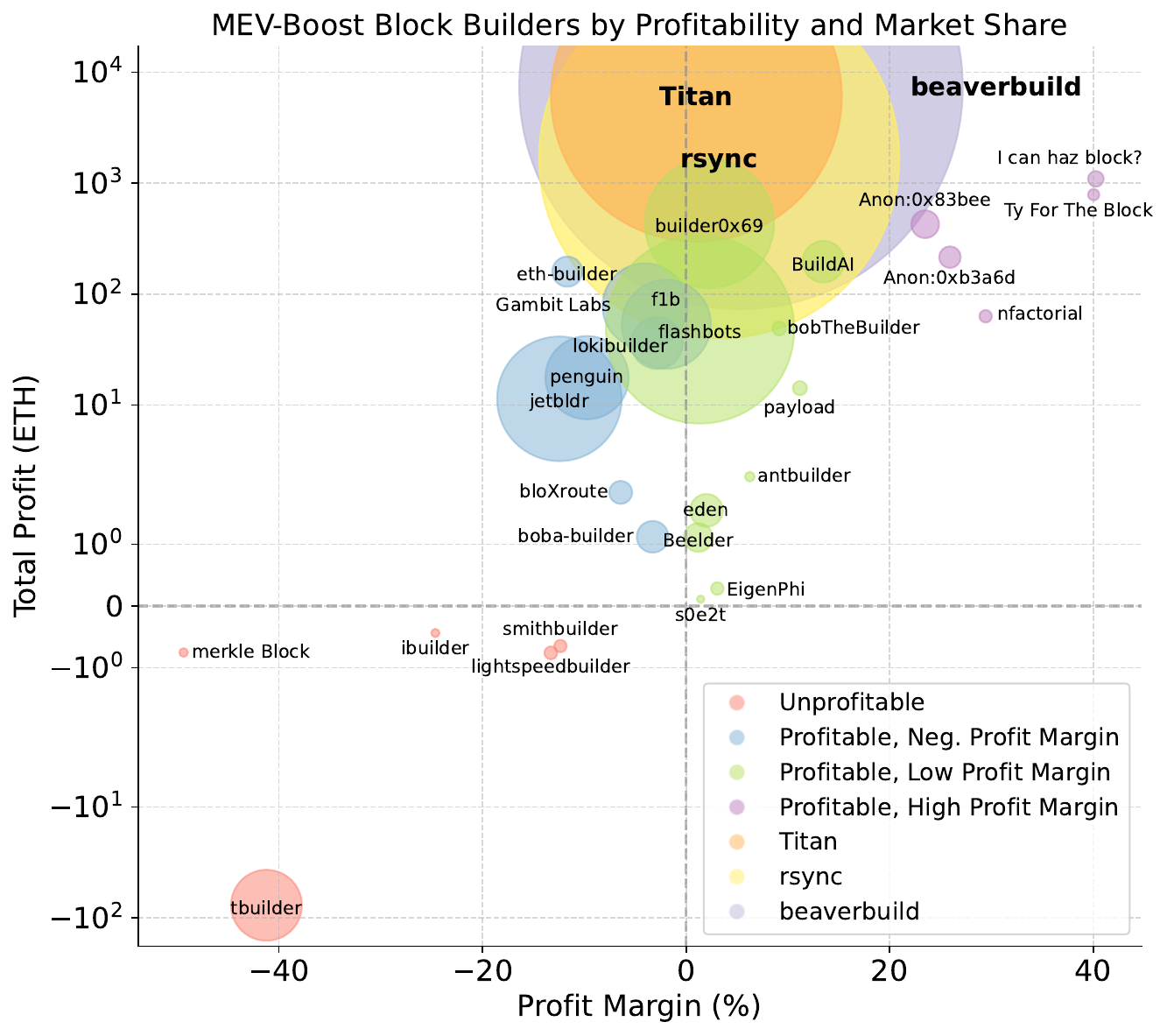}
    \caption{Bubble plot illustrating MEV-Boost builders' profit margin (average block value kept by the builder) on the x-axis and total profit in ETH on the y-axis. Bubble size represents the builder's market share, measured by the total number of blocks built. Color indicates profitability, with a \SI{15}{\%} profit margin threshold distinguishing high-profit margin builders. Builders with less than \SI{0.01}{\%} market share or a profit margin below \SI{-50}{\%} are omitted for brevity.}
    \label{fig:pm_vs_profits}  
    
\end{figure}

To preserve Ethereum's decentralized ethos and censorship resistance properties, fostering competition in the MEV-Boost block building auction is essential. As a first step, we must understand the competitive edges of the dominant players. MEV-Boost block builders currently have varying profitability levels, which do not strictly increase with the block market share they own, as shown in \Cref{fig:pm_vs_profits}. Thus, we need to study the critical factors for both obtaining market share and earning profits. Recent work \cite{yang_decentralization_2024} highlighted the pivotal role certain order flow providers play in winning blocks. We extend their insights by analyzing all active builders, order flow over time, and specific strategies to provide a comprehensive understanding of what drives builders' success in the MEV-Boost auction.\\


Our contributions can be summarized as follows:

\begin{itemize}
    \item We develop a novel methodology to identify key features in MEV-Boost builders’ success in winning blocks and earning profits, and provide metrics to monitor the builder market.

    \item We reveal that builders' block market share positively correlates with order flow diversity, while profitability correlates with access to order flow from \glspl{ep}, such as integrated searchers and external providers with exclusivity deals.


    \item We show a positive correlation between market share and profit margin among the top ten builders, with features such as exclusive signal, non-atomic arbitrages, and Telegram bot flow strongly correlating with both metrics.

    \item We highlight a ``chicken-and-egg'' problem where builders need differentiated order flow to profit, but only receive such flow if they have a significant market share.

    \item We discuss the implications of our findings for Ethereum block auctions and explore existing solutions for addressing the censorship threat posed by the currently centralized block builder market.

    
    
\end{itemize}

\section{Background and Related Work}
In this section, we present the necessary background and discuss the related work.

\subsection{Proposer-Builder Separation and the MEV-Boost Block Auction}\label{pbs-back}

The \gls{pbs} framework \cite{pbs} introduces the decoupling of block building and proposal tasks for staked Ethereum validators. While the validator remains responsible for signing and proposing the block, the new builder role handles the block content (i.e., execution payload) preparation. By delegating the \gls{mev} extraction task to builders, \gls{pbs} aims to lower the barrier for entry for validators, who no longer need to become proficient in block building.

Achieving the \gls{pbs} goal is essential for Ethereum's consensus security \cite{bahrani_centralization_2024}. However, it is challenging due to the difficulty of establishing a \textit{fair exchange} of value between the validator and the builder \cite{noauthor_fair_nodate}. Validators need assurance of payment and execution payload delivery by the builder, while builders must ensure their block content is protected against unbundling and is included on-chain. Although solutions to fair exchange problem have been discussed \cite{noauthor_why_2023, noauthor_epbs_nodate}, there is no enshrined \gls{pbs} implementation in the Ethereum protocol. 

Flashbots proposed an out-of-protocol \gls{pbs} implementation, MEV-Boost\cite{noauthor_flashbotsmev-boost_2024}, which introduces relay intermediaries to address the fair exchange issue. This implementation became active with the Merge \cite{noauthor_merge_nodate} in September 2022. In MEV-Boost, relays receive blocks from builders and validate\footnote{\textit{Optimistic} relays \cite{noauthor_optimistic-relay-documentationproposalmd_nodate} such as Ultra Sound Relay\cite{noauthor_ultra_nodate} are an exception. These relays can make the builder bids available to the proposer without validation since they have additional security mechanisms in place (e.g., collateralized builder funds).} them to ensure that the promised bid to the validator is paid and that the block is valid to become canonical. To protect builders against unbundling attacks, relays follow a commit-and-reveal scheme, disclosing only the block header when retrieving the proposer's signature. Although MEV-Boost requires trust in third parties, relays are presumed to act honestly due to their reputations. However, there have been instances where proposers exploited relay vulnerabilities to unpack and steal valuable \gls{mev} bundles from searchers \cite{noauthor_subverting_nodate, noauthor_equivocation_2023}, indicating that the fair exchange problem remains unresolved.

\begin{figure}[t!]
    \centering
    \includegraphics[width=0.9\linewidth]{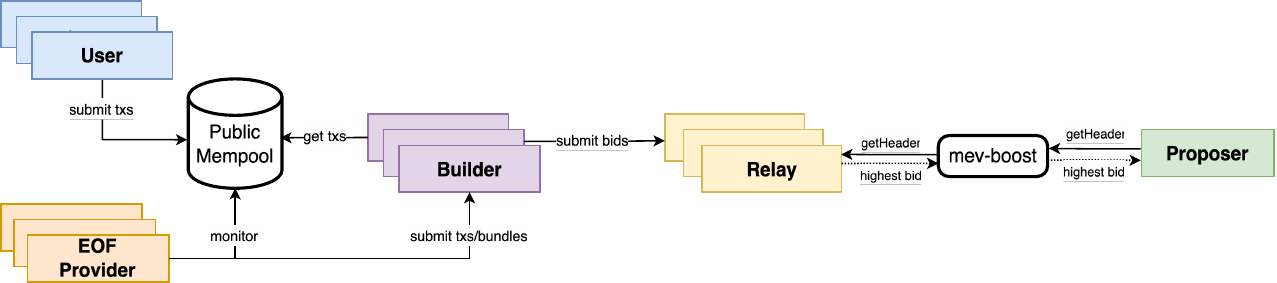}
    \caption{MEV-Boost block production process. Users (blue) submit transactions to the public mempool, accessible to every builder. EOF providers (orange), including \gls{mev} searchers and valuable order flow bots, monitor user transactions and bundle them with their own transactions or directly submit individual transactions to builders. Builders (purple) submit blocks with bids to the relays (yellow), which make them available to the proposer (green) through the mev-boost middleware.}
    \label{fig:pbs}  
\end{figure}

An MEV-Boost block auction round \cite{noauthor_mev-boost_proposal} follows the \SI{12}{-}second slot structure of the Ethereum consensus protocol \cite{noauthor_ethereumconsensus-specs_2024}. To win the right to produce the block proposed at slot $n$, builders start competing at slot $n-1$ by sending bids to the relays. Over time, builders increase the value of their bids, derived from the fees offered by users sending transactions to the public mempool and \gls{eof} providers (e.g., \gls{mev} searchers, Telegram bots \cite{banana, maestro}) submitting valuable bundles and transactions to the builders' private \gls{rpc} endpoints. Relays validate the builder bids and make them available to the proposer, who can continuously poll the \textit{getHeader} endpoint on the \textit{mev-boost middleware} to receive the highest bid from every registered relay. Eventually, the validator blindly signs the accepted bid's block header and submits it to the winner relay, who publishes the full block to the network. The described MEV-Boost process is summarized in \Cref{fig:pbs}.

\subsection{Maximal Extractable Value}
\gls{mev} refers to the sum of value extractable from a blockchain in any given state, through transaction ordering, insertion, and exclusion \cite{daian_flash_2020}. Using these techniques and analyzing the blockchain and network state, \textit{\gls{mev} searchers}, running bots and sophisticated smart contracts,\footnote{Example \gls{mev} bot contract: \texttt{0x6980a47bee930a4584b09ee79ebe46484fbdbdd0}} execute strategies such as \gls{dex}-\gls{dex} or \gls{cex}-\gls{dex} arbitrages, sandwiches, and liquidations to earn profits \cite{qin2022quantifying, heimbach_non-atomic_2024}.

Due to the competitive, time-sensitive,\footnote{Arbitrages between an off-chain \gls{cex} and an on-chain \gls{dex} or two \glspl{dex} on different blockchains can be considered time sensitive as execution is non-atomic.} and state-dependent nature of \gls{mev} extraction, searchers must pay fees like bribes to builders for prioritized inclusion and execution in blocks. As a result, builders earn a proportion of the \gls{mev} that the searchers extract. However, they must also share their earnings with the proposers to win the MEV-Boost block auction, as described previously (see \Cref{pbs-back}). Thus, \gls{mev}, originating from benign user transactions altering the blockchain state, flows through various entities in the block production pipeline and provides economic incentives for participants throughout the supply chain.

\subsection{Order Flow Auctions}
\glspl{ofa} are auctions where users share trade orders as unsigned transactions with searchers who compete in a sealed-bid format to backrun them. Users are incentivized by refunds up to \SI{90}{\%} of the value bidders extract from backrunning opportunities \cite{mev_blocker}, along with frontrun protection. \glspl{ofa}, such as MEVBlocker \cite{mev_blocker} and MEV-Share \cite{noauthor_mev-share_2024}, provide an \gls{rpc} endpoint for users to privately submit transactions instead of sending them to the public mempool. These transactions are braodcasted, either in full detail or selectively, to bidders who compete to submit the highest value bundle to block builders. Builders then refund \gls{ofa} users using the fee from the winning searcher's bundle.

\subsection{Related Work}
Previous work has focused on  Ethereum block building auctions from theoretical, game-theoretic, and empirical standpoints. \cite{builder_dominance} reveals that searchers prefer submitting bundles to builders with high-market share, and new entrants need to subsidize to obtain market share unless they operate their own searchers. \cite{bbp_thomas} conducts an empirical study examining the strategic behavior of MEV-Boost block builders, focusing on their bidding behavior and order flow strategies, and establishes core metrics for analyzing builder profiles and identifying their competitive edges. \cite{wu_strategic_2023} provides a game-theoretic model of the MEV-Boost auction and adopts an agent-based model to simulate builder strategies, showing the importance of latency and access to order flow. \cite{yang_decentralization_2024} identifies pivotal order flow providers for block builders and measures the competition and efficiency of the MEV-Boost auctions. \cite{gupta_centralizing_2023} shows that integrated \gls{hft} builders who extract top block position opportunities, such as \gls{cex}-\gls{dex} arbitrages, are favored to win the block auction when price volatility increases. \cite{heimbach_non-atomic_2024} introduces heuristics tailored to detect \gls{cex}-\gls{dex} arbitrages, confirming that price volatility and the arbitrage volume are correlated. The work provides empirical evidence that certain builders, such as \texttt{beaverbuild} and \texttt{rsync}, run their own searchers to extract this value. \cite{bahrani_centralization_2024} proves the necessity of a competitive \gls{pbs} auction to enable homogeneous proposer rewards and avoid concentration of stake among validators proficient in extracting \gls{mev} rewards. Block proposal timing games have been studied in \cite{schwarz-schilling_time_2023, oz_time_2023}, showing the marginal value of time for validators to earn more rewards from the builder bids. \cite{wahrstatter_time_2023} measures the MEV-Boost market concerning the entities involved in the block production pipeline, and \cite{pbs_promise} analyzes the Ethereum landscape with a comparison between \gls{pbs} and non-\gls{pbs} blocks.

\section{Methodology}\label{methodology}
This section outlines the methodology for identifying transaction labels, collecting empirical data, and calculating MEV-Boost block metrics.

\subsection{Transaction Labels Taxonomy}\label{taxo}
\begin{figure}[t!]
    \centering
    \includegraphics[width=0.9\linewidth]{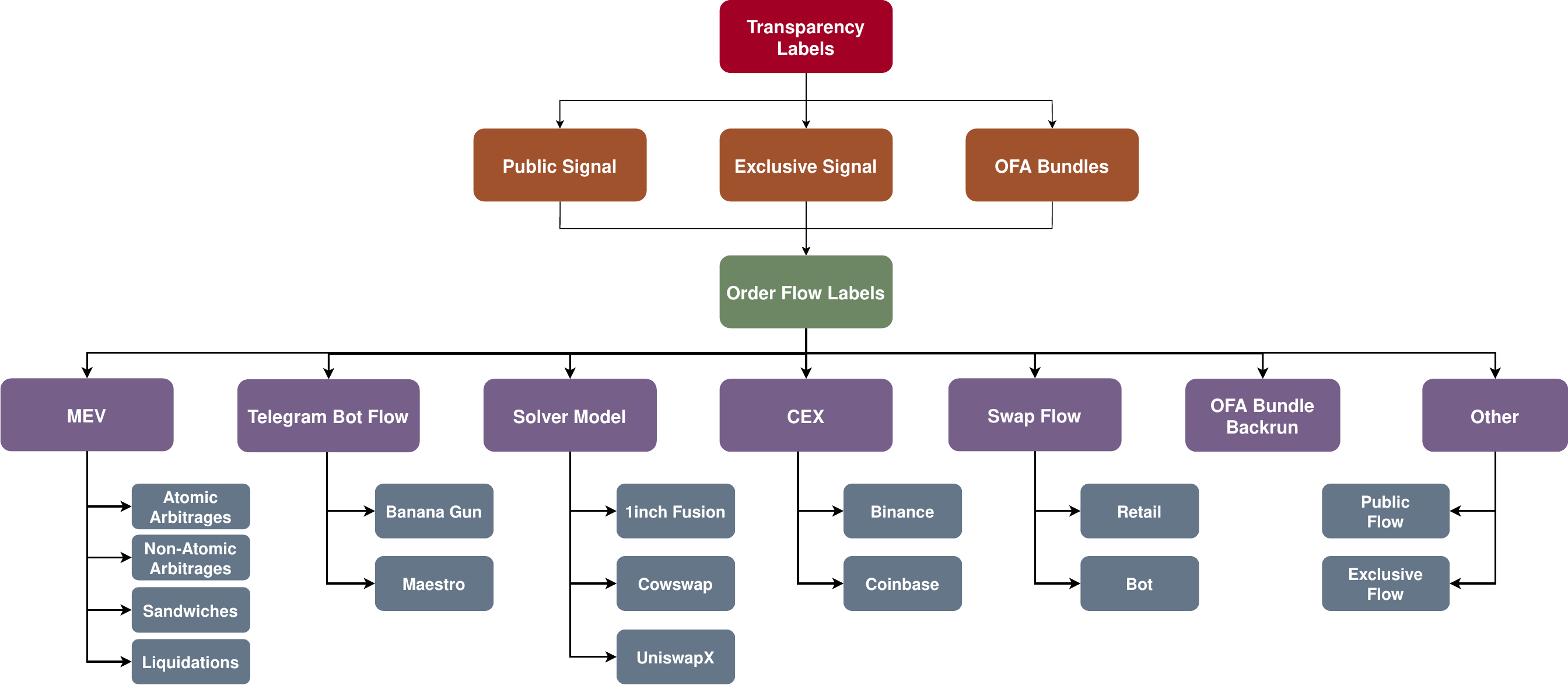}
    \caption{Taxonomy of transaction transparency and order flow labels. Every transaction has a transparency label (orange) based on its visibility on the network level and an order flow label (purple) determined by its objective. The gray labels stemming from the order flow labels represent the detailed categories we consider or popular providers of such order flow.}
    \label{fig:of_taxonomy}  
\end{figure}

We classify Ethereum transactions to understand the competitive edge different types provide to block builders based on the payments they offer. To this end, we determine a \textit{transparency} and an \textit{order flow} label for every transaction, following the taxonomy presented in \Cref{fig:of_taxonomy}.

\subsubsection{Transparency Labels}
Ethereum transactions submitted to public \gls{rpc} endpoints propagate through a peer-to-peer network and enter nodes' public mempool. These transactions are available to every builder running a node in the Ethereum network. However, some transactions are submitted exclusively to a builder's private \gls{rpc} endpoint, bypassing the public mempool. Additionally, certain transactions submitted to private endpoints are revealed to a selected group of entities, as occurs in \glspl{ofa}. As part of our methodology, we apply the following conditions to determine the transparency label of a transaction:


\begin{itemize}
\item \textit{Public Signal}: Recorded in the mempool of at least one monitored node.

\item \textit{Exclusive Signal}: Not recorded in the mempool of any monitored nodes, indicating exclusive submission to the block builder.

\item \textit{OFA Bundle}: Part of an \gls{ofa} bundle, either as the original user transaction or the searcher backrun. The heuristics for identifying \gls{ofa} bundles are detailed in \Cref{heuristics}.

\end{itemize}

\subsubsection{Order Flow Labels}\label{ofl}
Ethereum transactions have various objectives, ranging from simple ETH transfers to complex \gls{mev} strategies. Consequently, they have different valuations for their originators, reflected through the payments offered to block builders. To understand the impact of builders' access to order flow on their success in the MEV-Boost auction, it is critical to measure the value of different order flow types. Therefore, we assign an order flow label for every transaction included in our examined MEV-Boost blocks. While we consider detailed categories, the taxonomy is not exhaustive and omits certain known \gls{mev} strategies, such as \gls{jit} liquidity \cite{capponi_paradox_2024} and cross-chain arbitrages \cite{obadia_unity_2021,noauthor_cross-chain_2023}, and misses unknown long-tail strategies employed by the searchers.

We first determine if the transaction can be identified as a known \textit{\gls{mev}} type, such as an atomic \gls{dex}-\gls{dex} or a non-atomic \gls{cex}-\gls{dex} arbitrage, a sandwich frontrun or backrun,\footnote{Sandwiched user transactions are labeled according to their original objective.} or a liquidation. If no \gls{mev} labels are matched, we check if the transaction was submitted by a \textit{Telegram bot} such as Banana Gun \cite{banana} or Maestro \cite{maestro}. These bots create wallets, store private keys for users, and execute highly time-sensitive strategies on their behalf, such as token mint sniping or long-tail token trading, in exchange for fees. Next, we consider \textit{solver model} transactions submitted by solvers, fillers, and resolvers of protocols such as Cowswap\cite{noauthor_cow_nodate}, UniswapX\cite{uniswapX}, and 1inch Fusion \cite{noauthor_1inch_nodate}, fulfilling user limit orders. Additionally, as blockchain users require \textit{\glspl{cex}} like Binance \cite{noauthor_binance_nodate} or Coinbase \cite{noauthor_coinbase_nodate} to deposit and withdraw cryptocurrencies, we also detect and label such transactions individually. Further, if a transaction includes an ERC-20 token transfer or a swap but does not belong to prior categories, we label it as \textit{retail swap} or \textit{bot swap} depending on its initially interacted smart contract. If this contract is a known, non-\gls{mev} contract such as the Uniswap Universal Router,\footnote{Uniswap Universal Router contract address: \texttt{0x3fc91a3afd70395cd496c647d5a6cc9d4b2b7fad}} we label the transaction as retail swap, likely originated through the frontend of the respective protocol. Otherwise, we consider the transaction a bot swap, interacting with an unlabeled contract, potentially executing long-tail \gls{mev} strategies or known \gls{mev} strategies undetected by our datasets and heuristics. We treat \textit{\gls{ofa} bundle backrun} as a distinct category, although user transactions within \gls{ofa} bundles are not labeled individually. Finally, any remaining transaction is labeled as \textit{other public flow} or \textit{other exclusive flow} depending on its transparency type. The former category includes transactions like simple ETH transfers or batch submissions by rollup sequencers, while the latter covers transactions such as \gls{mev} bot contract deployments by searchers. The detailed identification methodology for each label is presented in \Cref{heuristics}.

\subsection{Data Collection}\label{datacollection}
We curate an empirical dataset covering a six-month period of MEV-Boost blocks produced between October 1, 2023, and March 31, 2024, totaling \SI{1190617}{} blocks. We primarily develop our data collection methods using external platforms such as Dune Analytics \cite{dune} and open source them for reproduceability by the community. We first build a compound query on Dune \cite{noauthor_dune_nodate}, utilizing various datasets available on the platform \cite{github_atomic_mev_arb, github_atomic_mev_sandwiches, github_atomic_mev_sandwiched, query_cex_label, query_telegram_bot, github_mempool_dumpster, methodology_orderflow}. Through this query, we identify transaction properties beyond the standard payload data. These properties include transactions' direct payments to builders' coinbase address, ERC-20 transfers, certain \gls{mev} activities, trade volumes and fees in USD, \gls{mev} searcher and solver labels, Telegram bot labels, router contract labels, and mempool visibility. We export the resulting dataset, involving \SI{174240225}{} labeled transactions. Additionally, we identify builder-controlled public keys using the extra data field of the Ethereum blocks \cite{danning_builder_keys}. Finally, we obtain the remaining necessary data for our methodology, such as all MEV-Boost payloads, bids on the UltraSound relay \cite{noauthor_ultra_nodate},\footnote{UltraSound relay was one of the most dominant relays during the course of this study \cite{wahrstatter_mev-boost_nodate}.} further builder public keys and searcher addresses, and other smart contract and transaction labels from external resources summarized in \Cref{tab:datasets} in \Cref{externaldata}.

\subsection{MEV-Boost Block Metrics}
The value of an MEV-Boost block is derived from the transactions in it. Each transaction offers a payment to the block builder, depending on its valuation for the originator. There are two ways a transaction can make an on-chain payment: through transaction fees and direct value transfers to the builder. Based on the total value builders receive from these transaction payments and their private valuation for the block, they offer a bid to the validator and make a payment if they win the auction. In this section, we present the methodology for calculating the true value, validator payment, and builder profit of MEV-Boost blocks.

\subsubsection{True Block Value}
We define the set of MEV-Boost blocks we analyze in our study as $B$. For each transaction $t \in T_b$, where $b \in B$ and $T_b$ represents all transactions included in $b$, we denote its gas priority fee (i.e., tip) and direct transfer to builder's coinbase address as $t_{tip}$ and $t_{coinbase}$, respectively. We refer to the refund amount of builder transactions in \gls{ofa} bundles as $t_{refund}$. For failed transactions included in $b$, we set their coinbase payment to 0 and only consider their tip payment to the builder. We define the true value ($TV$) of a block as:
\[b_{TV} = \sum_{t \in T_b} (t_{\text{tip}} + t_{\text{coinbase}} - t_{\text{refund}}).\]
We note that $TV$ is only a best-effort estimate based on available on-chain data. For a more robust measure, $TV$ calculation must account for private valuations of the searchers and builders and off-chain payment deals.

\subsubsection{Validator Payment}\label{validator_payment}
For a given MEV-Boost block $b$, we denote the validator payment ($VP$) as $b_{VP}$, referring to the amount tipped to the designated validator's proposer fee recipient address by $b$. Although most MEV-Boost blocks involve a builder transaction at the block's last index to make this payment, there are edge cases where the payment is made by an address related to the builder, or there is no payment from the builder, and the validator's proposer address is set as the block's fee recipient \cite{builder_profits} . We assume the first case makes no difference when calculating $b_{VP}$, but we omit the blocks from our analysis where the proposer is set as the fee recipient. While builders can still profits from such blocks if they have own searcher transactions or through off-chain deals, since there is no on-chain value transfer to the builder, we ignore them. This strategy might be used when the builder aims to gain market share while avoiding the basefee for the validator payment transaction, especially when the expected profit from the block is low. In \Cref{tab:validator_payment_patterns} in \Cref{validator_payment_appendix}, we summarize builders' excluded blocks.

In this paper, we calculate $b_{VP}$ by identifying the validator payment transaction. We avoid using the payload value reported by relays as $b_{VP}$ since this value can be manipulated if the relay calculates a builder's bid value by the balance difference of the proposer \cite{noauthor_blockscore_2022}, leading to a miscalculation of validator earnings. One reason for such discrepancies is stake withdrawals \cite{withdrawals}. When a builder includes a withdrawal transaction to the validator of the block, such relays over-report the bid by considering the withdrawal amount as part of the validator payment done by the builder, potentially causing profit loss for the validator \cite{noauthor_distortion_nodate}.  We share empirical results of builders' validator payment patterns in \Cref{tab:validator_payment_patterns} in \Cref{validator_payment_appendix}.

\subsubsection{Builder Profit}
We denote the builder profit ($BP$) of an MEV-Boost block $b$ as $b_{BP}$. When calculating $b_{BP}$, we handle the following edge cases:
\begin{itemize}
    \item Besides the validator payment and \gls{ofa} refund transactions, we do not deduct other builder value transfers from $b_{TV}$ when calculating $b_{BP}$. Since such transfers can be issued to builder-controlled \glspl{eoa} or smart contracts, deducting their value or using builder coinbase address balance change may underestimate $b_{BP}$ \cite{builder_profits},
    \item When the builder does not issue a validator payment transaction, we check if the last transaction of the block still pays the validator. If that is the case and the validator's proposer address is not the fee recipient of the block, we assume this payment is made on behalf of the builder, either by an address associated with the builder or one controlled by the UltraSound relay. The latter occurs when the builder opts to use UltraSound's recently deployed bid-adjustment feature \cite{ultrasound_bid}, which alters the builder's winning bid to be \SI{1}{wei} above the second-best bid available on any relay and refunds a proportion of the bid surplus. Before March 5, 2024, the entire difference was refunded, but since then, UltraSound has been taking half of it, which is eventually transferred by the builder. We refer to this relay payment ($RP$) value of block $b$ as $b_{RP}$. In \Cref{tab:validator_payment_patterns} in \Cref{validator_payment_appendix}, we present the total amount paid by each builder to UltraSound.
\end{itemize}

In this paper, we define the on-chain builder profits as $b_{BP} = b_{TV} - b_{VP} - b_{RP}$, and calculate the profit margin ($PM$) of a builder from $b$, referring to the share of true block value kept by the builder, with $b_{PM} = \frac{b_{BP}}{b_{TV}}$.

\subsection{Limitations}\label{limitations}
Our methodology has the following limitations:
\begin{enumerate}
    \item \textbf{Transparency Labels}: Accuracy is limited by the mempool dataset's coverage across node providers \cite{github_mempool_dumpster}. Determining transaction exclusivity across builders requires access to losing MEV-Boost bids' execution payload data, which the relays do not publicly disclose.
    \item \textbf{Order Flow Labels}: Potentially includes false labels due to our strict heuristics (see \Cref{heuristics}) and use of external data (see \Cref{tab:datasets} in \Cref{externaldata}).
    \item \textbf{Builder Profits}:
    \begin{itemize}
        \item Overestimated if the builder has off-chain deals with providers such as \gls{mev} searchers or Telegram bots, and pays them for accessing their order flow.
        \item Underestimated if the builder is vertically integrated with any \gls{mev} searcher, as we only consider on-chain profits made through priority fees and coinbase transfers.
    \end{itemize}
    \item \textbf{\gls{mev} Searcher Labels}: Not exhaustive as we cannot detect all the bots one searcher operates, obscuring the significance of a single searcher across multiple addresses.
    \item \textbf{Bidding Data}: Can be biased as we only consider bids on the UltraSound relay.
\end{enumerate}

\section{MEV-Boost Decomposition}\label{mevboostdecomposition}
We present our measurements, decomposing the MEV-Boost auction to identify potential features driving builder success. First, we examine the builder market to observe the dominant players in market share and profits. Next, we analyze the order flow in MEV-Boost blocks to determine the most valuable flows and which builders receive them. Finally, we explore various strategies builders adopt to gain a competitive advantage in the auction.

\subsection{Builder Market Structure}\label{bms}
Between October 2023 and March 2024, \SI{39}{} different block builder entities won the MEV-Boost auction, producing \SI{1190617}{} blocks. \gls{brt} contributed the most blocks, accounting for \SI{34.86}{\%}, \SI{22.98}{\%}, \SI{22.74}{\%} of the total, respectively. In approximately \SI{8}{\%} of MEV-Boost blocks, the builder set the fee recipient address as the proposer, with \texttt{Titan} responsible for \SI{92.86}{\%} of these cases (see \Cref{tab:validator_payment_patterns} in \Cref{validator_payment_appendix}). We suspect \texttt{Titan} adopts this strategy when the block's profits are insufficient to cover the basefee of the validator payment transaction. In the following analyses, we exclude these blocks for reasons discussed in \Cref{validator_payment}.

\begin{figure}[hbtp]
    \centering
    \begin{subfigure}[b]{0.32\linewidth}
        \includegraphics[width=\linewidth]{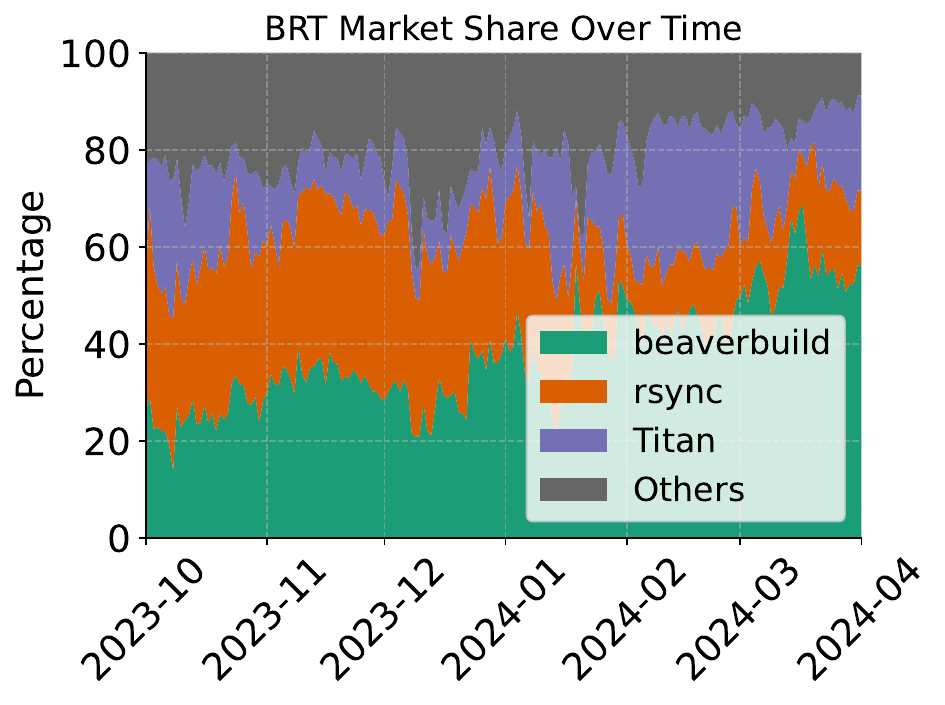}
        \caption{}\label{fig:market_share}
    \end{subfigure}
    \begin{subfigure}[b]{0.32\linewidth}
        \includegraphics[width=\linewidth]{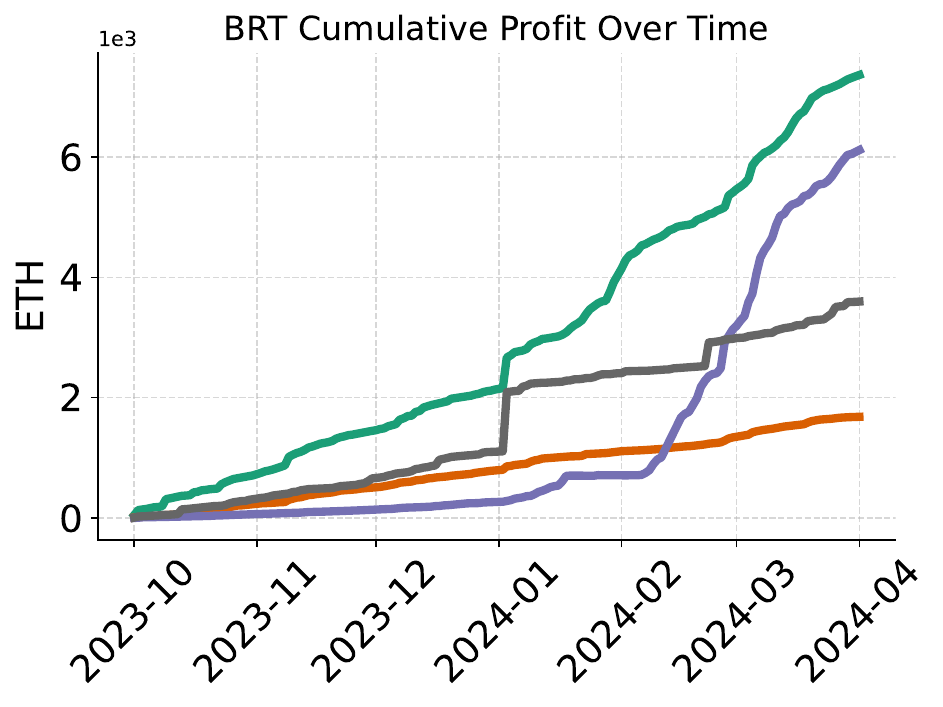}
        \caption{}\label{fig:cum_profits}
    \end{subfigure}
    \begin{subfigure}[b]{0.32\linewidth}
        \includegraphics[width=\linewidth]{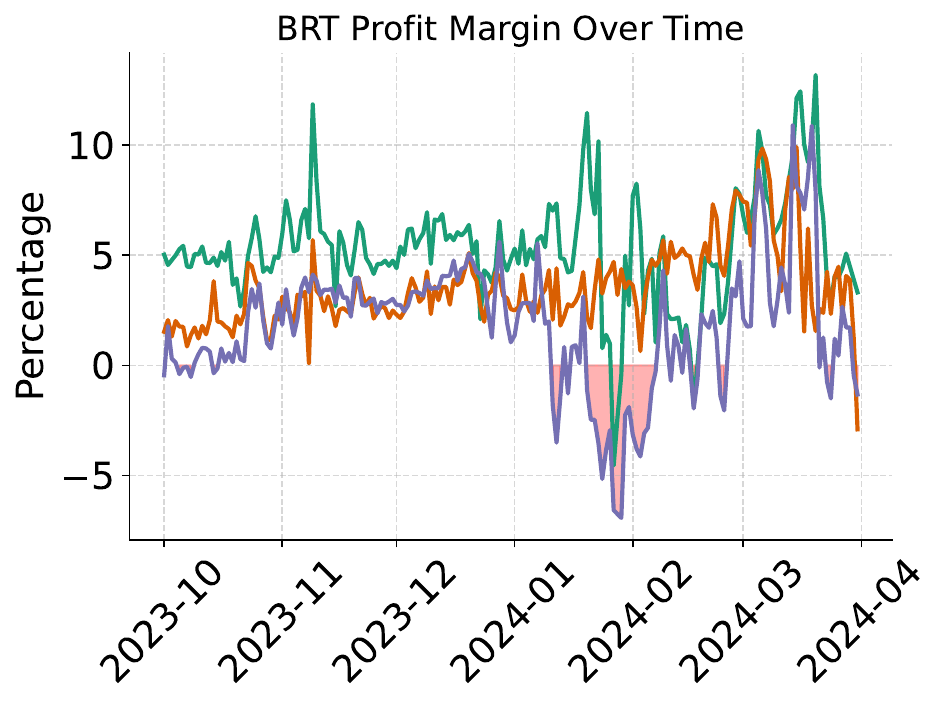}
        \caption{}\label{fig:pm_over_time}
    \end{subfigure}
     \caption{Trends in the market share, profits, and profit margin of the top three builders with highest market share, \gls{brt}, over time. \Cref{fig:market_share} is an area plot highlighting the changes in market share owned by \gls{brt} and the aggregated, remaining builders, denoted as ``Others'' in the legend. \Cref{fig:cum_profits} is a line plot showing the cumulative ETH profits of the same entities. \Cref{fig:pm_over_time} displays the daily profit margin changes of \gls{brt}. The area below 0 on the y-axis is filled with red to indicate days when the builder was, on average, unprofitable.}\label{fig:combined_market_profits}
\end{figure}

In \Cref{fig:combined_market_profits}, we show the market share (\Cref{fig:market_share}), cumulative profit (\Cref{fig:cum_profits}), and daily profit margin (\Cref{fig:pm_over_time}) of \gls{brt} over time. On average, these three builders construct \SI{80}{\%} of all blocks. We observe a surge in \texttt{Titan}'s market share and profits since February 2024, surpassing \texttt{rsync} in cumulative profits, approaching \texttt{beaverbuild}. Interestingly, when measured in USD, \texttt{Titan} has the highest total profit, reaching roughly \SI{19.7}{M} USD, despite building fewer blocks than \texttt{beaverbuild} and \texttt{rsync}, who earned \SI{19.4}{M} USD and \SI{4.27}{M} USD, respectively. Although the profits we calculate gives an insight about builders' earnings, the exact values can be different since we only consider the value provided through transaction priority fees and coinbase payments. Currently, we miss the upstream value builders make through their integrated searchers \cite{heimbach_non-atomic_2024,gupta_centralizing_2023}. Furthermore, we cannot track profits accurately if builders have any off-chain settlement with order flow providers.

Examining the builders besides \gls{brt}, we discover that only seven others have a market share nearly \SI{1}{\%} or more (see \Cref{tab:overall_stats}). Together, these top ten builders produced \SI{97.67}{\%} of the blocks but only earned \SI{83.85}{\%} of the \SI{52}{M} USD total builder profit. This discrepancy highlights the diverse builder profiles, with different specializations in gaining market share and profits.

\Cref{tab:overall_stats} summarizes the measurements for the top ten builders with the largest market share.\footnote{We only consider the MEV-Boost blocks where the builder is the fee recipient address.} The various profiles of the builders are reflected in their profitability metrics. \texttt{flashbots} and \texttt{builder0x69} behave neutrally, not winning blocks by paying more than they earn from the included transactions. In contrast, \texttt{jetbldr}, \texttt{penguin}, and \texttt{tbuilder} heavily subsidize and maintain a negative profit margin on average. However, except for \texttt{tbuilder}, remaining builders are overall profiting.\footnote{This can be due that subsidizing builders seldom earn significant profits, keeping them profitable.} \texttt{beaverbuild} has the highest profit margin at \SI{5.4}{\%} within the top ten, whereas across all builders, \texttt{I can haz block?} and \texttt{Ty For The Block} have approximately \SI{40}{\%} profit margins (see \Cref{tab:all_stats} in \Cref{all_overview}).

We note that \texttt{Titan} has a lower profit margin than \texttt{rsync}, despite earning significantly higher profits. While our profit calculation may not reflect precise values due to discussed issues, the discrepancy between profits and profit margin could be because \texttt{Titan} has more days with a negative profit margin compared to the other two top builders, who maintain a more consistent profit margin over time (see \Cref{fig:pm_over_time}). As \texttt{beaverbuild} and \texttt{rsync} run their own searchers \cite{gupta_centralizing_2023, heimbach_non-atomic_2024}, they receive a more consistent order flow stream. Conversely, \texttt{Titan} relies on external order flow providers who may not be as consistent, resulting in some blocks with significant profits and others that are entirely unprofitable. This pattern also supports our observation about the high percentage of \texttt{Titan} blocks where the proposer is set as the fee recipient (see \Cref{tab:validator_payment_patterns} in \Cref{validator_payment_appendix}).

\begin{table}[t!]
\centering
\footnotesize
\begin{tabular}{lcccccr}
\toprule
 & Total  & Market &  Total & Total & Total & Profit \\ 
Builder & Blocks & Share & Validator Payment & Subsidy & Profit & Margin \\ 
\ & [\#] & [\%] & [ETH] & [ETH] & [ETH] & [\%] \\ 
\midrule
beaverbuild & \SI{413868}{} & \SI{37.91}{} & \SI{49871.82}{} & \SI{-70.46}{} & \SI{7341.62}{} & \SI{5.4}{} \\ 
rsync & \SI{273126}{} & \SI{25.02}{} & \SI{33691.74}{} & \SI{-80.71}{} & \SI{1679.73}{} & \SI{3.25}{} \\ 
Titan & \SI{177915}{} & \SI{16.3}{} & \SI{22025.19}{} & \SI{-61.49}{} & \SI{6083.95}{} & \SI{1.02}{} \\ 
flashbots & \SI{74636}{} & \SI{6.84}{} & \SI{6918.79}{} & \SI{0.0}{} & \SI{48.22}{} & \SI{1.38}{} \\ 
builder0x69 & \SI{35083}{} & \SI{3.21}{} & \SI{3878.92}{} & \SI{0.0}{} & \SI{434.64}{} & \SI{2.3}{} \\ 
jetbldr & \SI{32670}{} & \SI{2.99}{} & \SI{1192.15}{} & \SI{-73.21}{} & \SI{11.41}{} & \SI{-12.45}{} \\ 
f1b & \SI{16716}{} & \SI{1.53}{} & \SI{1251.63}{} & \SI{-12.39}{} & \SI{53.59}{} & \SI{-1.94}{} \\ 
Gambit Labs & \SI{15281}{} & \SI{1.4}{} & \SI{889.45}{} & \SI{-19.75}{} & \SI{78.42}{} & \SI{-4.03}{} \\ 
penguin & \SI{14622}{} & \SI{1.34}{} & \SI{1028.28}{} & \SI{-41.77}{} & \SI{17.7}{} & \SI{-9.73}{} \\ 
tbuilder & \SI{10584}{} & \SI{0.97}{} & \SI{338.38}{} & \SI{-79.1}{} & \SI{-77.01}{} & \SI{-41.23}{} \\ 
\bottomrule
\end{tabular}
\caption{Market Share and Profitability Metrics of Top Ten Builders}\label{tab:overall_stats}
\end{table}

\subsection{Order Flow Breakdown}\label{generalof}
In this section, we analyze the order flow in the MEV-Boost blocks, focusing on their transparency and significance. We use the labels defined in \Cref{taxo}.

\subsubsection{Transparency}\label{transp}
We first examine the transparency of the Ethereum order flow. Recent work \cite{yang_decentralization_2024} showed an increasing share of value coming from \gls{eof}. We confirm their findings and examine the transparency of individual order flow labels and builders' blocks.

\begin{figure}[t!]
    \centering
    \begin{subfigure}[b]{0.45\linewidth}
        \includegraphics[width=\linewidth]{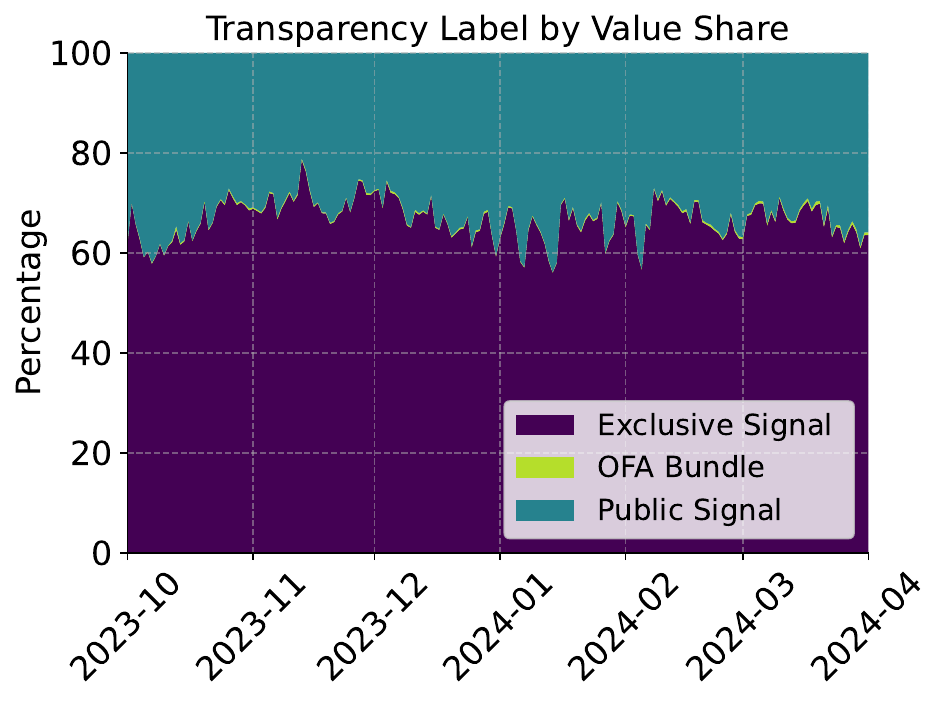}
        \caption{}\label{fig:transparency_value}
    \end{subfigure}
    \begin{subfigure}[b]{0.45\linewidth}
        \includegraphics[width=\linewidth]{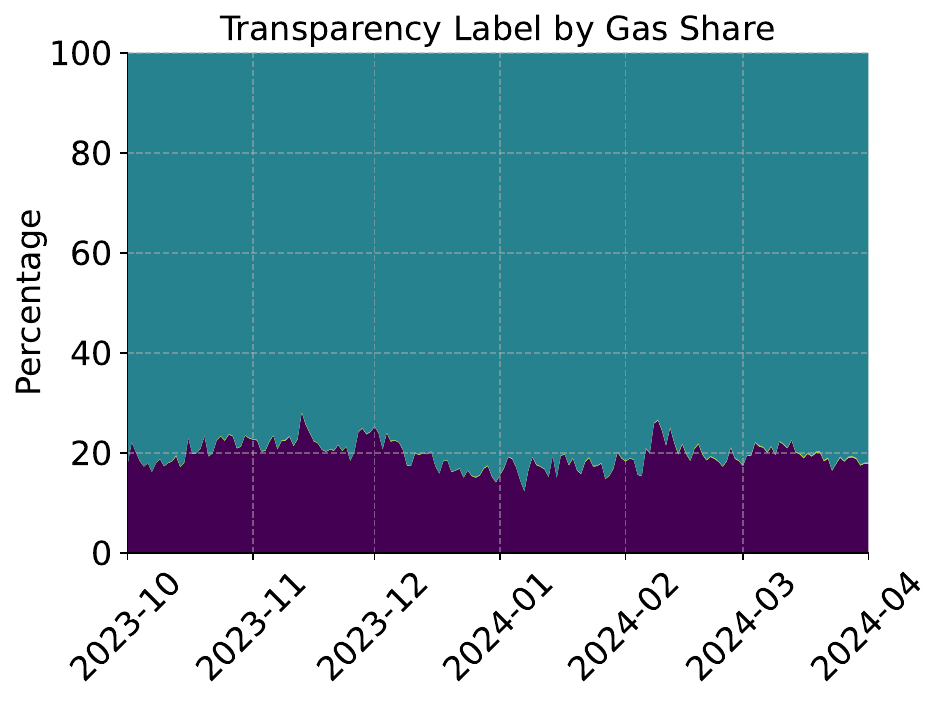}
        \caption{}\label{fig:transparency_gas}
    \end{subfigure}
    \vspace{10pt}
    \begin{subfigure}[b]{0.45\linewidth}
        \includegraphics[width=\linewidth]{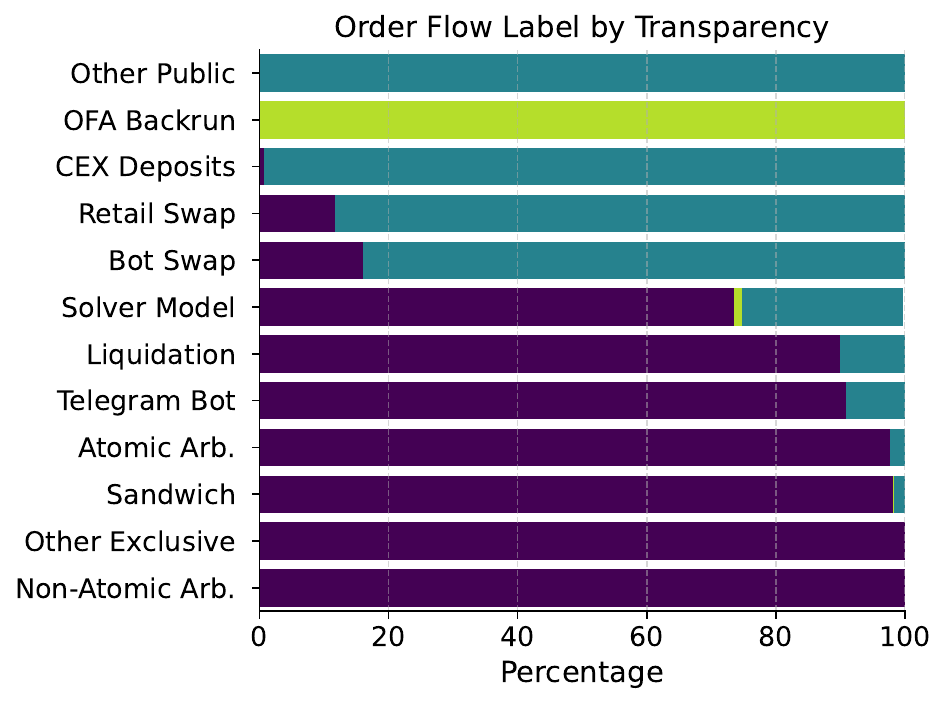}
        \caption{}\label{fig:of_transparency}
    \end{subfigure}
    \begin{subfigure}[b]{0.45\linewidth}
        \includegraphics[width=\linewidth]{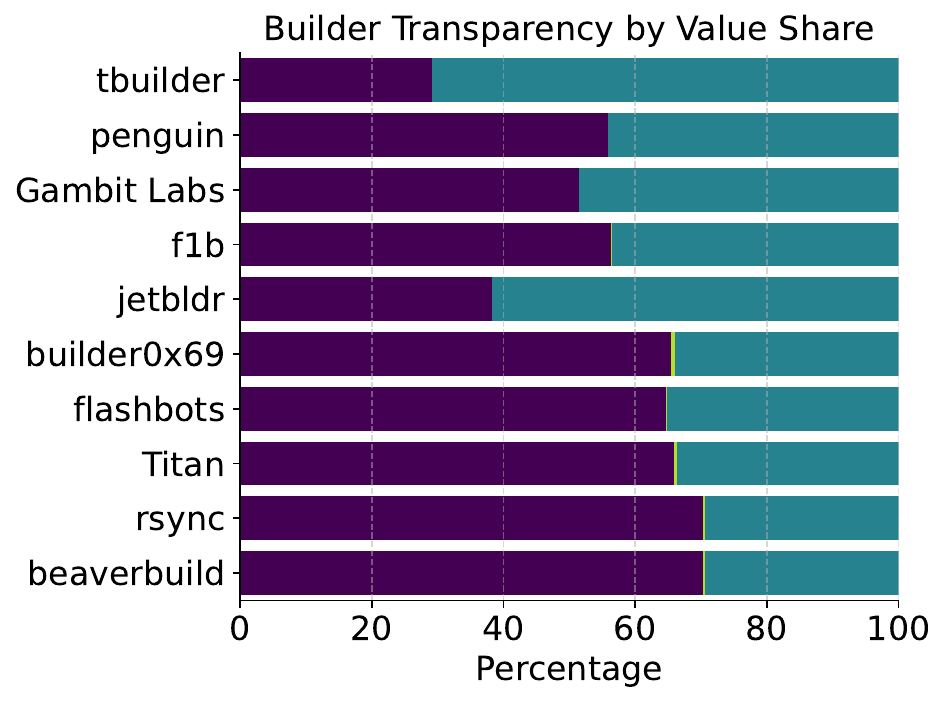}
        \caption{}\label{fig:builder_transparency}
    \end{subfigure}
    \caption{\Cref{fig:transparency_value} and \Cref{fig:transparency_gas} are area plots highlighting the share of value provided and blockspace consumed (in gas) by each transparency label over time. \Cref{fig:of_transparency} is a horizontal bar plot showing the transparency of each order flow label, measured in total transaction volume. \Cref{fig:builder_transparency} is a horizontal bar plot indicating the share of value that the top ten builders with the highest market share receive from each transparency label. Builders on the y-axis are ordered in ascending order based on their market share, with the builder with the highest share listed at the bottom. The legend in \Cref{fig:transparency_value} applies to all figures.}\label{fig:transparency}
\end{figure}

In \Cref{fig:transparency_value}, we show the share of true block value of each transparency label over time. Exclusive transactions and bundles, referred to as exclusive signal, provide \SI{66.69}{\%} of all value while consuming only \SI{19.6}{\%} of blockspace in terms of gas (see \Cref{fig:transparency_gas}). Similar to the results in \cite{yang_decentralization_2024}, we find that this flow constitutes \SI{71}{\%} of all block value in more than \SI{50}{\%} of the blocks, with nearly five times more value per unit of gas consumed than public signal. While almost all blocks have at least one exclusive and one public transaction, \gls{ofa} bundles occur in only \SI{4}{\%} of them, highlighting the scarce adoption of such protocols.\footnote{MEVBlocker and MEV-Share (through Flashbots Protect \cite{fbprotect}) have a larger volume if we consider the exclusive user transactions submitted to them which were not part of an \gls{ofa} bundle.} We summarize our measurements of transparency labels in \Cref{tab:of_stats} in \Cref{of_appendix}.

Next, we investigate the transparency of individual order flow labels, expecting certain types to avoid the public mempool to protect their valuable strategies. As shown in \Cref{fig:of_transparency}, every order flow label has a dominant way its involved in Ethereum blocks, corresponding to more than half of its total volume. Telegram bot flow and \gls{mev} strategies, including atomic and non-atomic arbitrages, sandwiches,\footnote{Sandwiches are mostly exclusive as we do not account the sandwiched user transactions in this label. Such transactions are labeled according to their original category, such as retail swap.} and liquidations, mostly bypass the public mempool. On average, \SI{99}{\%} of all \gls{mev} order flow is exclusively submitted to the builders. Conversely, retail and bot swap flows are primarily submitted to the public mempool, exceeding \SI{80}{\%} of their total volume. Interestingly, a relatively significant share of solver model flow is involved in \gls{ofa} bundles (\SI{1.28}{\%}), receiving refunds while fulfilling user trade orders. Notably, Cowswap solver model is known to be forwarding its flow to the MEVBlocker \gls{ofa} protocol.

Finally, we assess the transparency of builders' blocks. In \Cref{fig:builder_transparency}, we present the share of total value the top ten builders with the largest market share receive from public and exclusive signal and \gls{ofa} bundles. \gls{brt}, with the largest market share and most profits, receive the highest relative value from \gls{eof}, exceeding \SI{65}{\%} of their total. The builders with the lowest profit margins among the top ten, \texttt{jetbldr} and \texttt{tbuilder}, receive the least value from \gls{eof}, with shares below \SI{40}{\%}. Among builders outside the top ten who have greater than \SI{0.01}{\%} market share, more extreme concentrations are observed, with exclusive value shares ranging between \SI{1}{\%} and \SI{91}{\%} (see \Cref{fig:all_builder_transparancy} in \Cref{detailed_of_breakdown}).

\subsubsection{Significance}
\begin{figure}[t!]
    \centering
    \begin{subfigure}[b]{0.59\linewidth}
        \includegraphics[width=\linewidth]{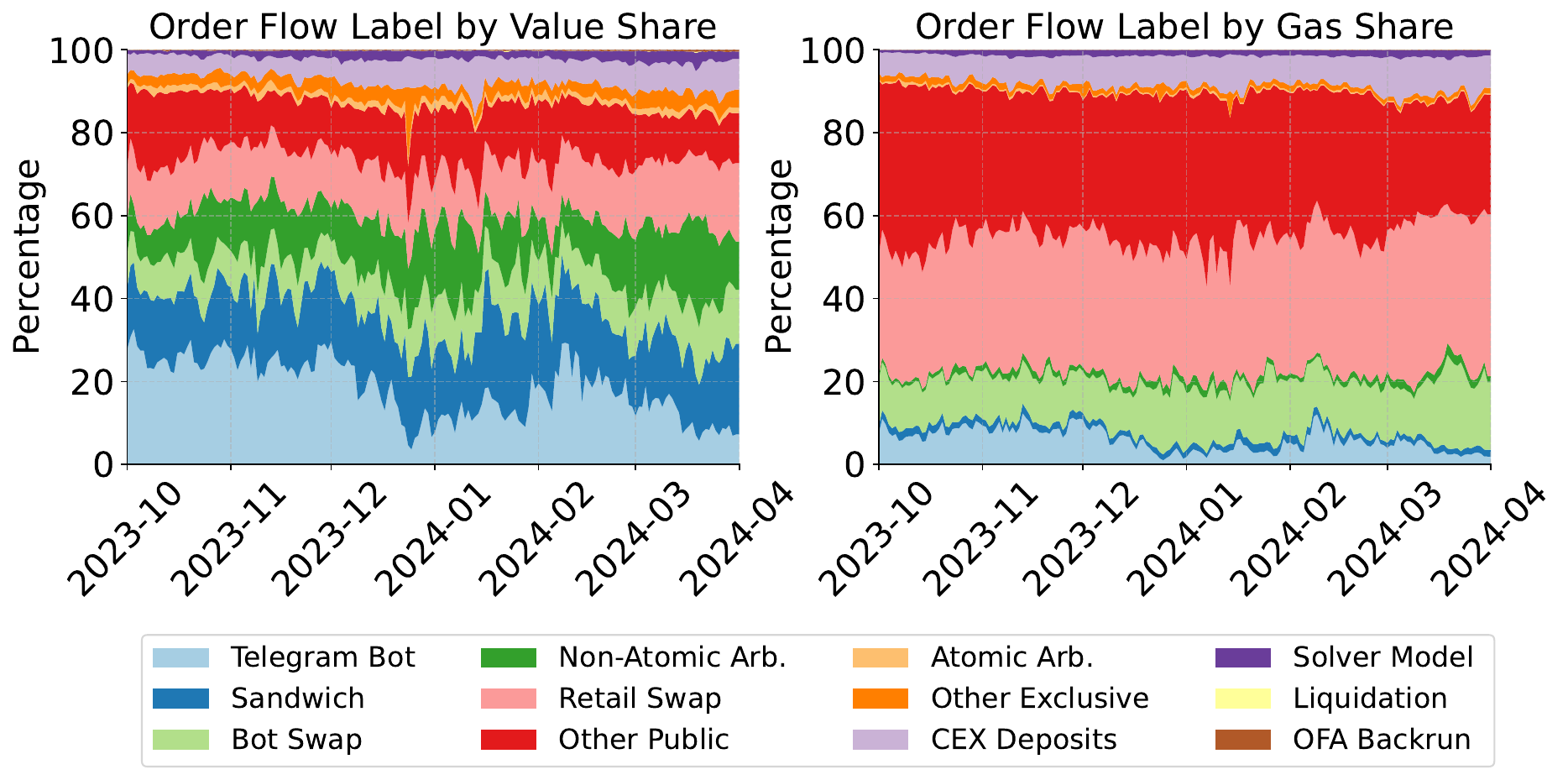}
         \caption{}\label{fig:of_prominence}
    \end{subfigure}
    \begin{subfigure}[b]{0.40\linewidth}
        \includegraphics[width=\linewidth]{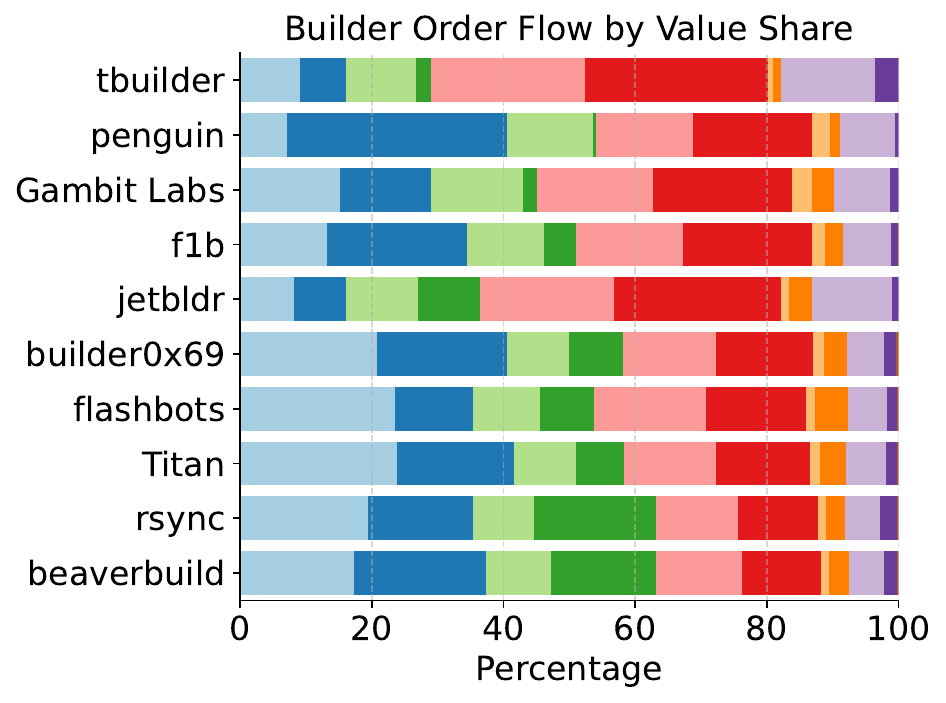}
     \caption{}\label{fig:builder_of_composition}
    \end{subfigure}
   
     \caption{\Cref{fig:of_prominence} shows the significance of order flow labels over time. The left and right panels depict the share of total block value provided and blockspace consumed (in gas) by each order flow label. \Cref{fig:builder_of_composition} displays horizontal bars representing the share of value the top ten builders with highest market share receive from each order flow label. Builders on the y-axis are ordered in ascending order based on their market share, with the builder with the highest share listed at the bottom. Order flow labels in \Cref{fig:builder_of_composition} use the same coloring as indicated in the legend in \Cref{fig:of_prominence}.}\label{fig:of}
\end{figure}

The order flow included in a block is a good estimator of its true value. To understand valuable flows, we analyze MEV-Boost blocks for the significance of the order flow labels. We find that Telegram bot and \gls{mev} order flows (see \Cref{ofl}) cumulatively contribute around \SI{51}{\%} of all block value while consuming merely \SI{10}{\%} of all gas spent in MEV-Boost blocks (see \Cref{fig:of_prominence}). These flows provide around \SI{53}{\%} of all block value in more than half of the blocks. Notably, they are primarily exclusively submitted to block builders, as shown in \Cref{fig:of_transparency}. In contrast, order flow labels such as retail swap and other public flow, most frequently observed in the public mempool, cumulatively provide a little over a quarter of the total block value while consuming more than \SI{67}{\%} of the total blockspace. These measurements of order flow labels, also summarized in \Cref{tab:of_stats} in \Cref{of_appendix}, indicate that valuable order flow is predominantly exclusively submitted, and blocks are mostly filled with less valuable, public transactions.

One of the drivers of competition in the MEV-Boost auction is the order flow builders receive \cite{wu_strategic_2023}. As order flow labels have varying levels of public and exclusive volume (see \Cref{fig:of_transparency}) and value (see \Cref{tab:of_stats} in \Cref{of_appendix}), we must analyze builders' order flow compositions concerning these labels to understand who possesses the valuable flow.


In \Cref{fig:builder_of_composition}, we show the order flow composition of the top ten builders with the highest market shares. Builders have varying compositions, with certain builders receiving a more significant share from specific order flows than others. Some of the most valuable flows, such as Telegram bot flow and non-atomic arbitrages, are more significant in the compositions of the top three builders, \gls{brt}, while almost all builders receive a considerable share of sandwich flow, which is a common strategy among \gls{mev} searchers. Furthermore, we examine the order flow composition of builders outside the top ten and discover more remarkable concentrations, with a single order flow providing more than \SI{80}{\%} of the total value a builder receives, such as the sandwich flow of \texttt{s0e2t}, as shown in \Cref{fig:all_builder_of} in \Cref{detailed_of_breakdown}.

\subsection{Builder Strategies}\label{builderstrats}
MEV-Boost block builders adopt various strategies to gain a competitive edge. While order flow makes up their value, bidding and latency strategies also play an important role during the auction process \cite{wu_strategic_2023}. In this section, we measure the significance of four builder strategies.

\subsubsection{Block Packing}
Authors in \cite{gupta_centralizing_2023} discuss the importance of accessing value from top block positions to win the block auction. They show through a theoretical model that builders who earn more from \gls{tob} are likely to dominate the \gls{pbs} auction. We analyze the block packing strategies of builders to support this outcome with empirical data. To this extent, we measure the share of the value builders earn from \gls{tob}, \gls{bob}, and \gls{eob} positions. We define \gls{tob} as the first \SI{10}{\%} of transactions based on their normalized index in the block, following \cite{ankit_chiplunkar_ankitchiplunkar_why_2023}. Similarly, \gls{eob}, denotes the bottom \SI{10}{\%}. Lastly, \gls{bob} represents the middle \SI{80}{\%} of the block. Builder transactions are excluded to prevent skewing the data.

In \Cref{fig:position}, we show the share of value the top ten builders by market share earn from transactions in each position group. \gls{tob} contributes the most significant value to all top builders except \texttt{penguin}, and \gls{brt} receive more than \SI{74}{\%} of their block value from order flow in \gls{tob}. The significance of \gls{tob} can be attributed to the fact that more than half of the total \gls{eof} volume resides in this position, providing, on average, \SI{88.5}{\%} of the total \gls{eof} value. Conversely, \gls{bob} and \gls{eob} only contribute approximately \SI{10.5}{\%} and \SI{1}{\%} of the total \gls{eof} value. Interestingly, \texttt{Titan}, \texttt{Gambit Labs}, and \texttt{penguin} receive a relatively considerable value from \gls{eob}. These builders could be allowing \gls{mev} searching at the end of their blocks, on the state transitioned to by the included transactions, and receiving significant payment from the granted searchers. The valuable block positions of the remaining builders are summarized in \Cref{tab:all_stats} in \Cref{all_overview}.

\subsubsection{Subsidization}
Builders who have just entered the market, not receiving significant volumes of valuable order flow, are expected to subsidize their blocks unless they run their own \gls{mev} searchers \cite{builder_dominance}. Since we previously found that certain builders have negative profits margins (see \Cref{tab:overall_stats}), we are interested in how significantly they subsidize to the win the auction. \Cref{fig:profits} shows the shares of blocks the top ten builders with highest market shares profited, subsidized, or were neutral, making dust amount or zero profit.\footnote{We set \SI{0.001}{ETH} as the dust amount threshold.} Notably, these measurements are solely based on on-chain profits and may not reflect the complete picture, as builders can profit from their integrated searchers or return a percentage of the value to order flow providers. 

The top three builders in market share, \gls{brt}, also have the highest shares of profitable blocks, with \texttt{beaverbuild} exceeding \SI{40}{\%}. While \texttt{Titan} is neutral in approximately \SI{50}{\%} of their blocks, this share goes up to \SI{68}{\%} when considering the blocks where they set the proposer as the fee recipient as zero profit as well (see \Cref{tab:validator_payment_patterns} in \Cref{validator_payment_appendix}).\footnote{We assume zero on-chain builder profit from such blocks although the builder can still earn profits from their integrated searchers or off-chain deals, as discussed in \Cref{validator_payment}.} \texttt{flashbots} and \texttt{builder0x69} predominantly break even, supporting the measurements summarized in \Cref{tab:overall_stats}. Conversely, \texttt{jetbldr}, \texttt{penguin}, and \texttt{tbuilder} win more than \SI{75}{\%} of their blocks by subsidizing, paying extra value on top of the on-chain value they receive from their order flow. Among builders outside the top ten, some adopt a profit-only strategy. Examples include \texttt{Anon:0x83bee}, \texttt{Anon:0xb3a6d}, \texttt{I can haz block?}, and \texttt{Ty For The Block}, who were also labeled as profitable, high-profit margin builders in \Cref{fig:pm_vs_profits}. The subsidization and profitability metrics of the rest of the builders are available in \Cref{tab:all_stats} in \Cref{all_overview}.

\subsubsection{Exclusive Order Flow Access}\label{eofproviders}
\gls{eof} contributes the majority of value to MEV-Boost blocks, as shown by our study (see \Cref{transp}) and previous works \cite{bbp_thomas,yang_decentralization_2024}. We also discovered that builders have varying level of \gls{eof} in their blocks (see \Cref{fig:builder_transparency}), indicating a non-uniform access. Extending studies on integrated builders and prominent order flow providers \cite{builder_dominance, gupta_centralizing_2023, searcherbuilder, heimbach_non-atomic_2024, yang_decentralization_2024}, we examine the significance of \gls{eof} providers for each builder. We consider providers including, public entities such as Flashbots Protect\footnote{Flashbots Protect is considered as the provider of the flow received by Flashbots's private \gls{rpc} endpoints, including MEV-Share \gls{ofa} bundles.} and MEVBlocker, \gls{mev} searchers, and solvers.\footnote{Solvers refer to the liquidity routing entity in the solver model introduced in \cref{taxo}.}

\begin{figure}[t!]
    \centering
    \begin{subfigure}[b]{0.32\linewidth}
        \includegraphics[width=\linewidth]{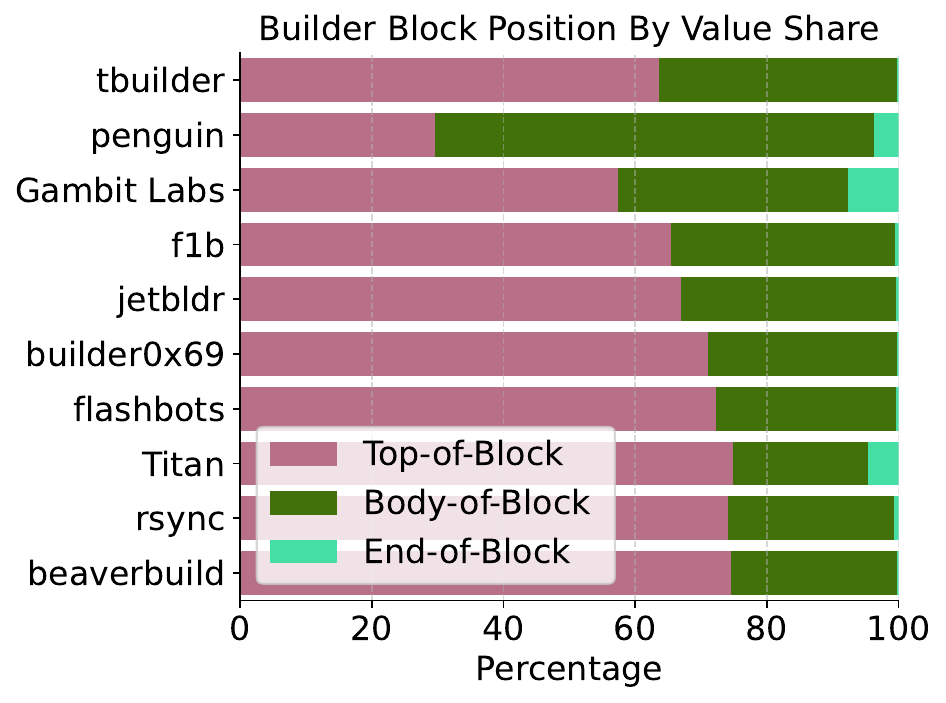}
         \caption{}\label{fig:position}
    \end{subfigure}
    \begin{subfigure}[b]{0.32\linewidth}
        \includegraphics[width=\linewidth]{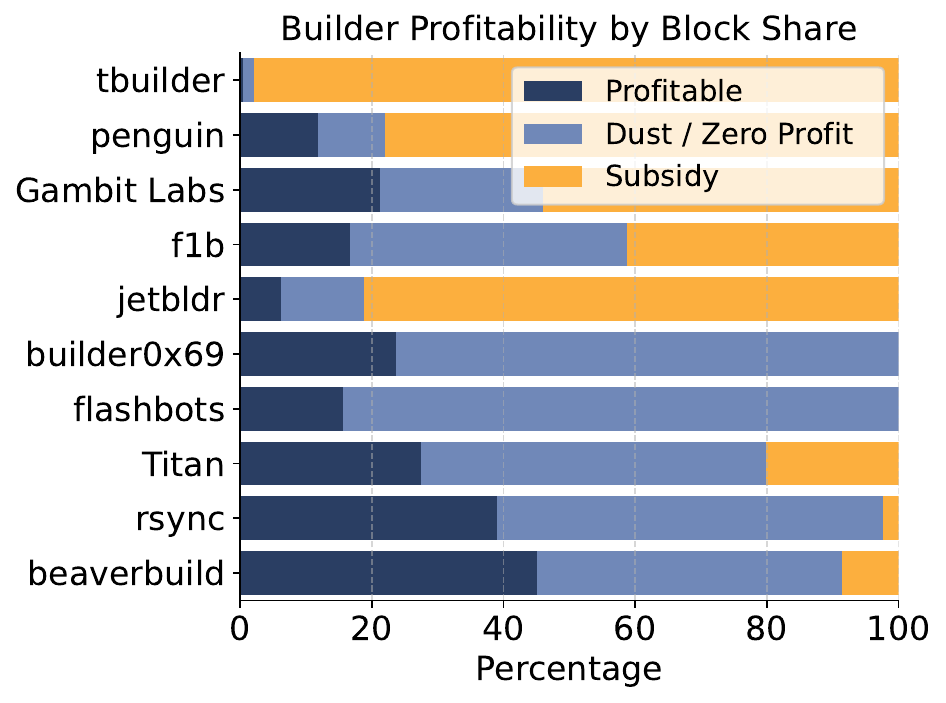}
     \caption{}\label{fig:profits}
    \end{subfigure}
    \begin{subfigure}[b]{0.32\linewidth}
        \includegraphics[width=\linewidth]{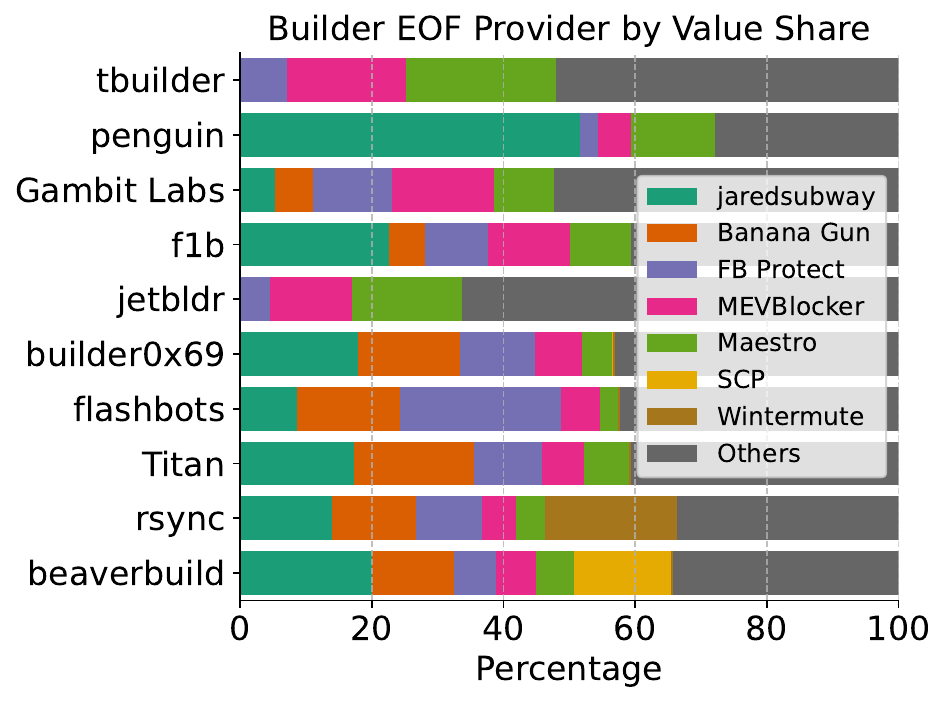}
     \caption{}\label{fig:eof_builders}
    \end{subfigure}
     \caption{\Cref{fig:position} is a horizontal bar plot showing the share of value the top ten builders by market share earn from \gls{tob}, \gls{bob}, and \gls{eob}. \Cref{fig:profits} is a horizontal bar plot presenting the share of blocks the top ten builders profited, subsidized, or made dust amount or zero profit. \Cref{fig:eof_builders} is a horizontal bar plot displaying the share of \gls{eof} value the top ten builders receive from the seven most significant \gls{eof} providers based on total value. In the plot legend, ``jaredsubway'' and ``FB Protect'' stand for  ``jaredfromsubway.eth'' and ``Flashbots Protect'' providers, respectively. The total \gls{eof} value of the remaining providers is aggregated and denoted as ``Others''. In all figures, builders on the y-axis are ordered from top to bottom in ascending order of market share.} \label{fig:strats}
\end{figure}

\Cref{fig:eof_builders} presents the shares of \gls{eof} value that the top ten builders with the highest market share receive from the seven most significant providers based on total value. These providers contribute roughly \SI{70}{\%} of all \gls{eof} value in more than half of the MEV-Boost blocks, and supply over \SI{48}{\%} of \gls{eof} value to each builder in the top ten, except \texttt{jetbldr}, who primarly sources \gls{eof} from other providers. We find that Flashbots Protect and MEVBlocker appear consistently, likely due to their public \gls{eof} access requirements, without needing private deals \cite{yang_decentralization_2024}. Maestro Telegram bot provides considerable flow to each top builder, unlike Banana Gun, which sends almost no flow to \texttt{jetbldr}, \texttt{penguin}, and \texttt{tbuilder} but provides significant value to \texttt{Titan}, especially starting from February 2024 (see \Cref{fig:titan} in \Cref{all_eof}). Additionally, \texttt{jetbldr} and \texttt{tbuilder} receive negligible \gls{eof} from the well-known sandwich searcher bot, jaredfromsubway.eth, while \texttt{penguin} gets more than \SI{50}{\%} of its \gls{eof} value from this provider, potentially explaining the prominence of sandwich flow in their order flow composition (see \cref{fig:builder_of_composition}). Finally, SCP and Wintermute are highly significant only for \texttt{beaverbuild} and \texttt{rsync}, respectively, as these providers are operated by them \cite{gupta_centralizing_2023, heimbach_non-atomic_2024}.\footnote{PLM and Rizzolver solvers, identified to be belonging to \texttt{beaverbuild} and \texttt{rsync} \cite{methodology_orderflow}, are aggregated with SCP and Wintermute providers, respectively.} Further integrations and private deals could exist among the remaining builders and \gls{eof} providers, as shown in \Cref{fig:all_eof_value_ratio} in \Cref{all_eof}.

\subsubsection{Latency and Bidding}\label{latency}
MEV-Boost block builders compete in latency when placing bids on the relays. Although the proposer is expected to release the block at slot start \cite{noauthor_ethereumconsensus-specs_2024}, the exact time is unknown. Due to the stochastic nature of this process, low latency is beneficial for builders 
 to update bids \cite{bbp_thomas}, reflecting changing valuations for the block over time and reacting to other bids.

Using data from the UltraSound relay, we examine MEV-Boost builders' latency and bidding behavior. We measure their bidding frequency by the average number of bids they place in each slot. Furthermore, we calculate the lag between their bid updates to understand how quickly they can reflect their new valuation and react to other bids. Lastly, we identify cases where builders strategically lowered their bid values through cancellations \cite{cancel}, a strategy discussed to be especially useful for builders running \gls{cex}-\gls{dex} arbitrage bots \cite{bbp_thomas}.

In \Cref{tab:latency_profile}, we summarize the bidding behavior of the top ten builders in market share for the blocks they won on UltraSound. We observe that the market share gap between \texttt{rsync} and \texttt{Titan} became more significant compared to their market shares when considering blocks from all relays (see \Cref{tab:overall_stats}), with a jump from approximately \SI{8.7}{\%} to \SI{12.2}{\%}. This could be because the UltraSound relay allows optimistic submissions \cite{noauthor_optimistic-relay-documentationproposalmd_nodate}, potentially favoring integrated builders such as \texttt{beaverbuild} and \texttt{rysnc} \cite{gupta_centralizing_2023, heimbach_non-atomic_2024}, as they can more frequently update their bids based on the value they derive from their searchers.

We measure that the top ten builders have diverse bidding strategies, with the average number of placed bids ranging from \SI{6.1} by \texttt{tbuilder} to \SI{49.64} by \texttt{penguin}, who also has the lowest average update lag between all of their bids in a slot (see \Cref{tab:latency_profile}). Builders can update bids quickly and frequently by having low latency relative to the relay's geographical position or by submitting new bids simultaneously from multiple public keys they control. 

\gls{brt}, who won the most auctions on the UltraSound relay, issue the highest number of cancellations, with, on average, more than one cancellation per block.\footnote{The average number of cancellations \gls{brt} issue per block increases from \SI{1.85}{} to \SI{2.09}{} after the introduction of UltraSound's bid adjustment feature \cite{ultrasound_bid}.} We note that, two of these builders, \texttt{beaverbuild} and \texttt{rysnc}, have relatively high non-atomic arbitrage shares (see \Cref{fig:builder_of_composition}) and were identified to be running their own \gls{cex}-\gls{dex} arbitrage searchers \cite{gupta_centralizing_2023,heimbach_non-atomic_2024}. Therefore, we confirm findings in \cite{bbp_thomas} regarding the use of cancellations by integrated builders. We also observe that some builders outside the top ten, such as \texttt{antbuilder}, who also has a relatively large non-atomic arbitrage flow (see \Cref{fig:all_builder_of} in \Cref{detailed_of_breakdown}), issued around ten cancellations per block they built (see \Cref{tab:all_latency} in \Cref{all_latency}).

We note the distribution of average winning times for builders, referring to the average slot time a builder's winning blocks were selected by the proposer. While the winning time for each builder is beyond the slot start (0ms), possibly due to timing games \cite{schwarz-schilling_time_2023,oz_time_2023}, there is approximately a \SI{580}{ms} difference between the earliest and the latest winner times. This difference can be related to the slot times builders start and stop bidding in the auction \cite{bbp_thomas}.

\section{Results}\label{results}
In this section, we identify the features associated with gaining market share and earning profits in the MEV-Boost auction. We apply a Spearman correlation with a p-value set to \SI{0.05}{} to assess the strength and direction of the associations based on rank orders. Spearman is used instead of Pearson to reduce sensitivity to outliers, which could be an issue due to the diverse set of block builders in our study. We exclude builders with less than \SI{0.01}{\%} market share, focusing on the remaining \SI{33}{} builders (see \Cref{tab:all_stats} in \Cref{all_overview}).

\begin{table}[t!]
\centering
\footnotesize
\begin{tabular}{lcccccr}
\toprule
\ & Total & Avg. & Avg. & Avg. & Total & Avg. \\ 
Builder & Blocks  & Bids  & Update Lag  & Winner Time  & Cancels & Cancels  \\ 
\ & [\#] & [\#] & [ms]  & [ms] & [\#] & [\#] \\ 
\midrule
beaverbuild & \SI{137721}{} & \SI{31.54}{} & \SI{130.41}{} & \SI{586.59}{} & \SI{248195}{} & \SI{1.74}{} \\ 
rsync & \SI{95310}{} & \SI{25.27}{}  & \SI{166.21}{} & \SI{531.38}{} & \SI{212504}{} & \SI{2.29}{} \\ 
Titan & \SI{47539}{} & \SI{31.11}{}  & \SI{186.25}{} & \SI{605.83}{} & \SI{91337}{} & \SI{1.78}{} \\ 
flashbots & \SI{32612}{} & \SI{16.25}{}  & \SI{411.36}{} & \SI{348.45}{} & \SI{3014}{} & \SI{0.1}{} \\ 
builder0x69 & \SI{23294}{} & \SI{20.64}{}  & \SI{183.08}{} & \SI{575.09}{} & \SI{17445}{} & \SI{0.65}{} \\ 
jetbldr & \SI{12960}{} & \SI{35.2}{} & \SI{103.08}{} & \SI{605.75}{} & \SI{904}{} & \SI{0.09}{} \\ 
Gambit Labs & \SI{9530}{} & \SI{15.85}{}  & \SI{349.05}{} & \SI{496.3}{} & \SI{492}{} & \SI{0.1}{} \\ 
f1b & \SI{9242}{} & \SI{25.42}{} & \SI{214.18}{} & \SI{680.45}{} & \SI{10494}{} & \SI{0.98}{} \\ 
penguin & \SI{7783}{} & \SI{49.64}{} & \SI{69.92}{} & \SI{652.91}{} & \SI{966}{} & \SI{0.13}{} \\ 
tbuilder & \SI{4786}{} & \SI{6.1}{}  & \SI{476.68}{} & \SI{104.71}{} & \SI{81}{} & \SI{0.01}{} \\ 
\bottomrule
\end{tabular}
\caption{Bidding and Latency Metrics of Top Ten Builders (UltraSound Relay)}
\label{tab:latency_profile}
\end{table}

\subsection{Order Flow Diversity}
The diversity of order flow a builder receives can indicate access to multiple sources, making the builder more competitive in auctions compared to those who depend on a specific order flow type, which may not always be available. To capture this, we measure the Shannon entropy \cite{shannon} of builders' order flow. With twelve unique order flow labels (see \Cref{ofl}), the Shannon entropy is defined as $H(X) = - \sum_{i=1}^{12} p_i \log_2(p_i)$, where $p_i$ represents the value share of each label in a builder's order flow composition. Higher entropy values indicate a more diverse order flow, while smaller values signify a concentrated composition.

\begin{figure}[t!]
    \centering
    \includegraphics[width=0.85\linewidth]{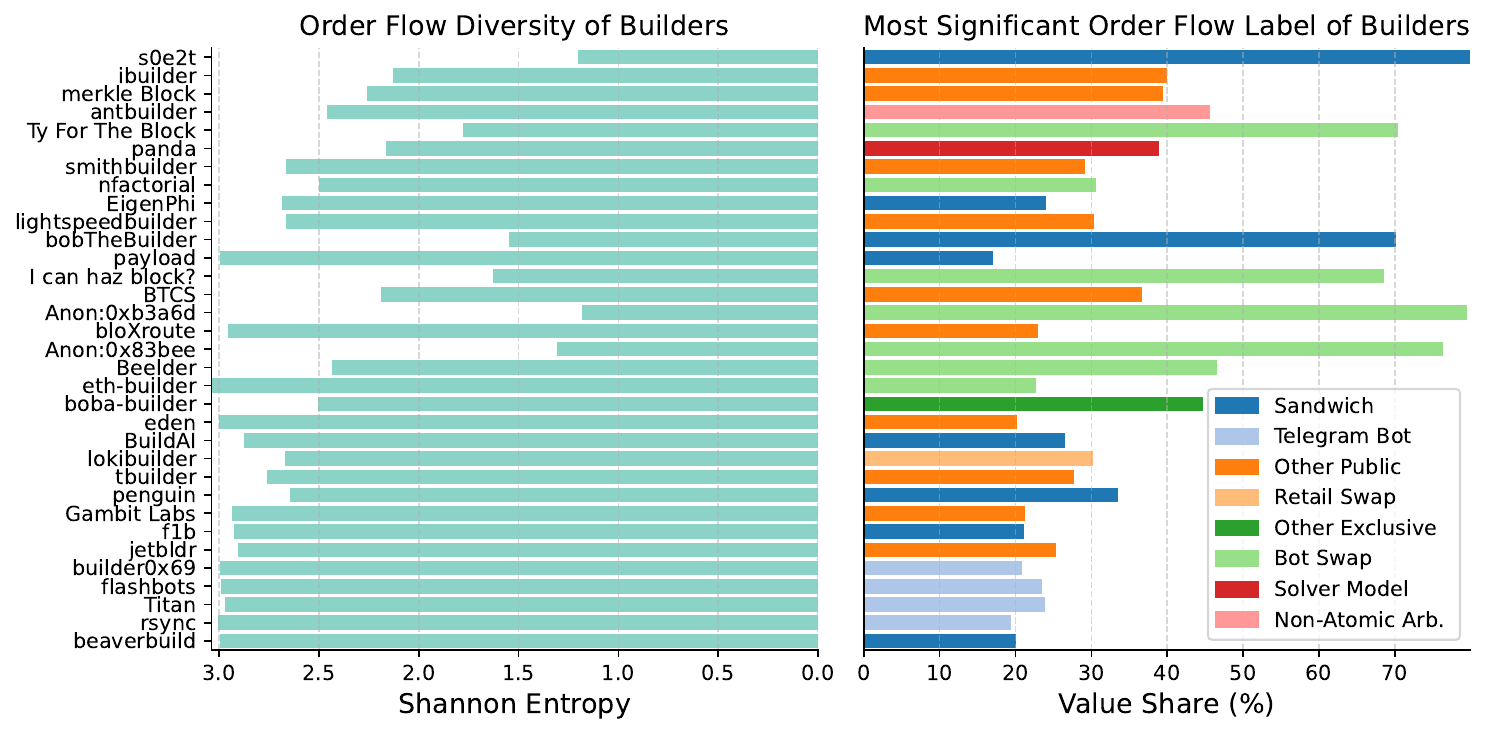}
    \caption{The left panel is a horizontal bar plot presenting each builder's order flow diversity through a Shannon entropy value. Builders are ordered from top to bottom in ascending order of market share. The right panel is a horizontal bar plot depicting each builder's \gls{msof}, referring to their highest value order flow label (in percentage). Bars are colored based on the order flow label.}
    \label{fig:of_entropy}
\end{figure}

In the left panel of \Cref{fig:of_entropy}, we show the entropy value of each builder's order flow composition. The top ten builders with the most market share have entropy values closer to the maximum ($H_{\text{max}} = \log_2(12) \approx 3.585$). However, among builders with smaller market shares, lower values are observed, suggesting a concentration around a few order flow labels. To better understand the order flow concentration, we measure builders' highest value share label, denoted as \gls{msof}. The outcome in the right panel of \Cref{fig:of_entropy} shows that the top ten builders have various \glspl{msof} labels, although all have less than \SI{32}{\%} value share. In contrast, across the remaining builders, \gls{msof} value share can exceed \SI{70}{\%}, indicating a highly concentrated order flow composition.

To examine the link between builders' success in winning the auction and the diversity of their order flow, we measure the correlation between builders' market share and order flow entropy. We find a significant positive correlation, with a coefficient of \SI{0.66}{} and a p-value of $2.22e-05$. This suggests that builders receiving different order flow types have an advantage in the auction and can win more frequently, as they are not dependent on a single type of flow. Conversely, builders with a concentrated order flow composition, where \gls{msof} is more prominent, can only be competitive when that specific order flow type is available.

\subsection{Exclusive Providers}\label{sp}
MEV-Boost block builders receive \gls{eof} from various providers, as discussed in \Cref{eofproviders}. Builders with own providers, such as \texttt{beaverbuild} and \texttt{rsync} \cite{gupta_centralizing_2023, heimbach_non-atomic_2024}, can produce profitable blocks as they access order flow not shared with others. To identify such exclusive relationships, we employ \gls{lda}, classifying providers that can successfully separate a builder's blocks from others. We consider the model's \gls{da} to be significant if it exceeds a threshold calculated using a binomial cumulative distribution \cite{combrisson_exceeding_2015}.

\begin{figure}[t!]
\centering
\includegraphics[width=0.85\linewidth]{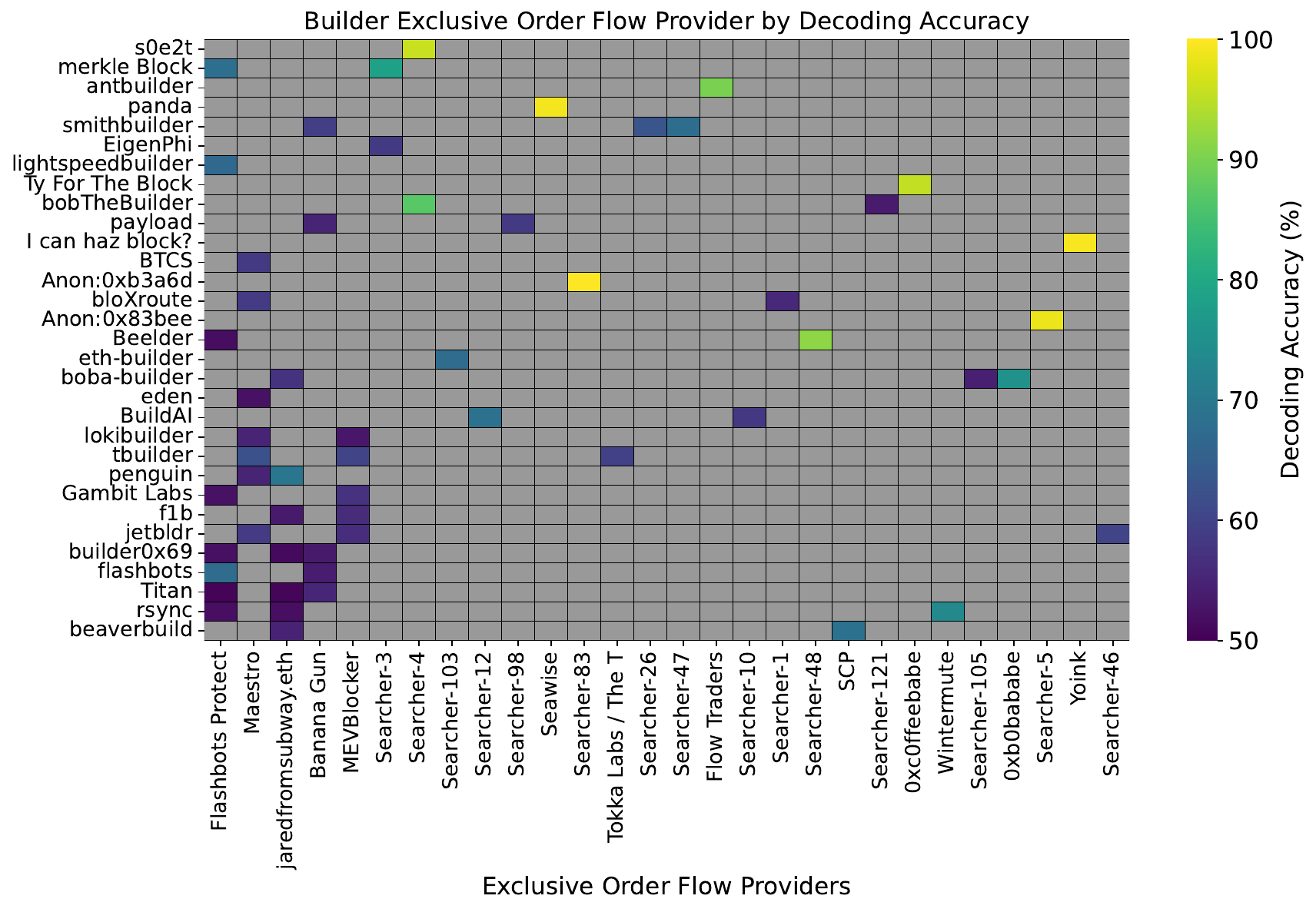}
\caption{Heatmap displaying the \gls{da} of \gls{eof} providers (x-axis) for each builder (y-axis). Colored cells indicate a statistically significant \gls{da}, whereas gray cells show negligible values. Columns with a single colored cell indicate \glspl{ep}.}
\label{fig:eof_heatmap}
\end{figure}

In \Cref{fig:eof_heatmap}, we present a heatmap showing the \gls{da} of \gls{eof} providers for each builder with at least one statistically significant provider. Unidentified providers are enumerated according to the mapping in \cite{eof_addresses}. A high \gls{da} indicates that the builder is distinguished by the significance of \gls{eof} they receive from the provider. If an \gls{eof} provider is prominent for many builders, it can be considered a neutral provider, accessible by all. Conversely, an \gls{eof} provider will be significant for only a single builder if they are vertically integrated or have an exclusivity deal. We define these providers as \glspl{ep}, identifiable by columns with a single colored cell in the heatmap.

We find that approximately \SI{55}{\%} of the builders have at least one \gls{ep}. These builders have a higher rate of profitable blocks (\SI{46}{\%}) compared to those without any \gls{ep} (\SI{20}{\%}). This result is supported by a Chi-square test on the number of profitable blocks built by the two sets of builders (Chi-square = \SI{33077.29}{}, p $<\SI{0.05}{}$). Additionally, we find a positive correlation between the rate of profitable blocks and whether builders have an \gls{ep}, with a coefficient of \SI{0.4}{} and a p-value of \SI{0.02}{}. 

Our results suggests that to produce profitable blocks, builders need exclusive relationships with providers who only supply significant \gls{eof} to them. We note that the correlation we measured could have been stronger if \gls{ep} profits were accounted towards builders, as certain \glspl{ep}, such as integrated searchers, do not necessarily reflect all the value they extract in their builder payments (see \Cref{limitations}). Interestingly, around \SI{46}{\%} of the builders with a negative profit margin (see \Cref{tab:all_stats} in \Cref{all_overview}) have an \gls{ep}, indicating that the value they subsidize on-chain could already be covered by the profits of their provider.

\subsection{Key Features for Market Share and Profitability}\label{key_features}

\begin{figure}[t!]
    \centering
    \begin{subfigure}[b]{0.49\linewidth}
        \includegraphics[width=\linewidth]{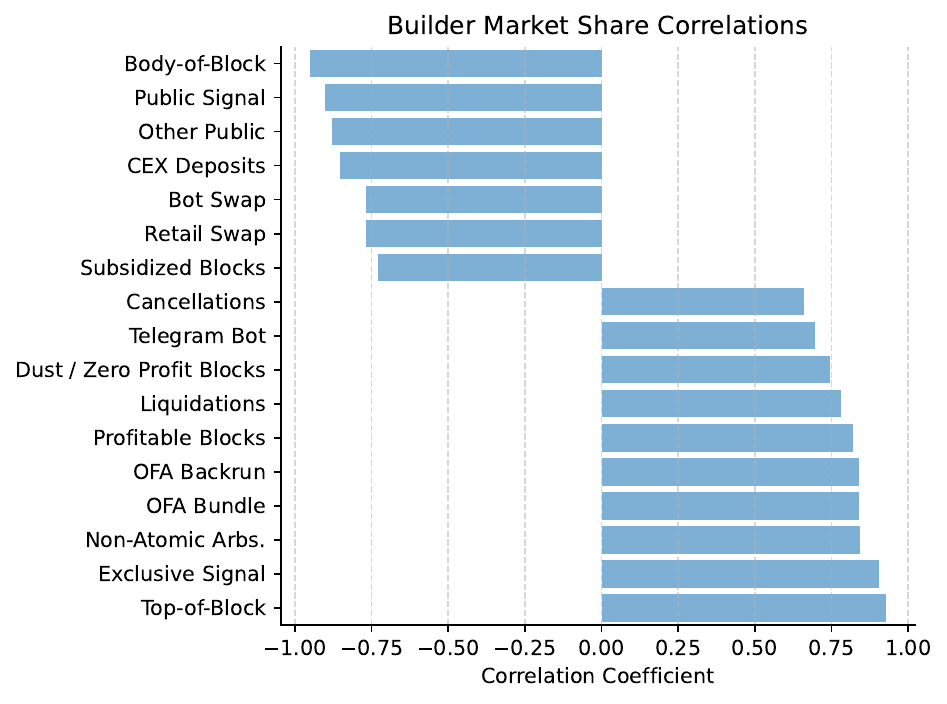}
         \caption{}\label{fig:corr_ms}
    \end{subfigure}
    \begin{subfigure}[b]{0.49\linewidth}
        \includegraphics[width=\linewidth]{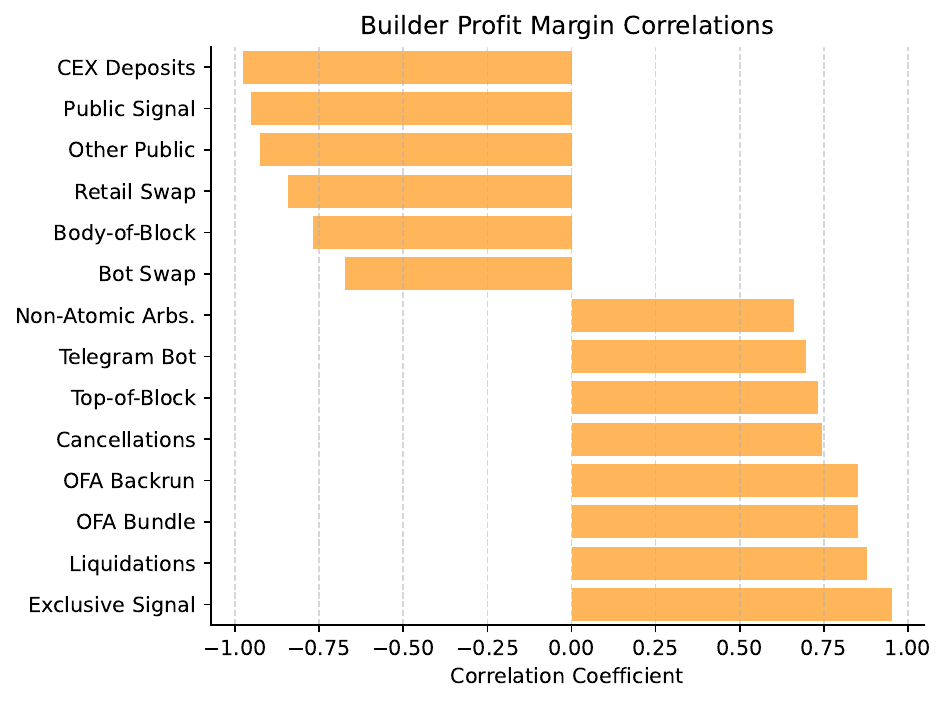}
     \caption{}\label{fig:corr_pm}
    \end{subfigure}
     \caption{Spearman rank correlations of order flow and builder strategy features for the top ten builders with the highest market share. \Cref{fig:corr_ms} is a horizontal bar plot showing the significant correlations between features and builders' market share. \Cref{fig:corr_pm} is a horizontal bar plot presenting the significant correlations between features and builders' profit margin.} \label{fig:corr}
\end{figure}

The MEV-Boost decomposition revealed various features potentially associated with builders' success in winning blocks and earning profits (see \Cref{mevboostdecomposition}). We find that, for the top ten builders who produced the most blocks, there is a significant positive correlation between their market share and profit margin (coef = \SI{0.87}{}, p $<\SI{0.05}{}$). Additionally, we identify overlapping features correlated with both metrics (see \Cref{fig:corr}), suggesting that the skills needed to succeed in both areas of the auction are similar.

Exclusive signal, non-atomic arbitrages, and Telegram bot flow strongly positively correlate with both market share (see \Cref{fig:corr_ms}) and profit margin (see \Cref{fig:corr_pm}), whereas public signal, including retail swaps, negatively correlates with them. This suggests that high-market share builders receive the more valuable exclusive flows, enabling them to keep significant profits, while those with smaller shares must fill their blocks with less valuable public transactions. For similar reasons, we observe a positive correlation of top ten builders' market share with their profitable block rate,\footnote{Correlation between profit margin and profitable block rate is omitted as they are related by definition.} and a negative correlation with their subsidy rate. Thus, we encounter the ``chicken-and-egg'' problem described in \cite{builder_dominance}: \textit{builders cannot profit from their blocks if they do not have order flow differentiating them from others, and they do not receive such flow unless they have a significant market share.}

Furthermore, top ten builders' success in gaining market share is strongly positively correlated with the value they derive from order flow in \gls{tob}, supporting the findings in \cite{gupta_centralizing_2023} (see \Cref{fig:corr_ms}). Conversely, \gls{bob} has the strongest negative correlation. Similar correlations are observed with builders' profit margin (see \Cref{fig:corr_pm}). We expect that valuable \gls{eof} received by successful builders is prioritized for execution at \gls{tob}, allowing them to operate on the latest transitioned state and avoid unexpected changes caused by prior transactions.

Finally, we find that builders' market share and profit margin are positively correlated with the number of bid cancellations they place per block they win on the UltraSound relay, as shown in \Cref{fig:corr}. This highlights the importance of such bidding strategies during the auction process, allowing builders to lower their active bid, either to avoid making a loss due changing private valuation for the block or to keep higher profits if there is a significant gap with the second highest bid. This is also facilitated by the bid adjustment feature of UltraSound \cite{ultrasound_bid}. Interestingly, we do not find a significant correlation with bid update lag, suggesting that latency is not necessarily linked with succeeding in the auction. However, this result may be biased as our latency analysis only considers the blocks relayed through UltraSound. For definitive results, bidding data on other relays must be examined as well.

\section{Discussion}
The current \gls{pbs} implementation, MEV-Boost, has successfully enabled uniform access to \gls{mev} rewards for Ethereum validators, thereby avoiding centralization of the consensus participants. However, the current design of the block building auction promotes competition in \gls{eof} access and latency, raising the barrier to entry for new builders and leading to a centralized market (see \Cref{results}). While this centralization is not as detrimental to the protocol's security as a centralized validator set, it still undermines Ethereum's censorship resistance \cite{fox_censorship_2023, wahrstatter_blockchain_2023} and \textit{neutrality} \cite{noauthor_uncrowdable_2024} properties.

MEV-Boost block builders are incentivized to access valuable order flow from diverse sources and, notably, from distinguished \glspl{ep} (see \Cref{results}). The top two builders with the highest market shares, \texttt{beaverbuild} and \texttt{rsync}, are operated by \gls{hft} firms running integrated searchers specializing in non-atomic arbitrages \cite{gupta_centralizing_2023, heimbach_non-atomic_2024}. The third best builder, \texttt{Titan}, appears to have an exclusive deal with the Banana Gun Telegram bot starting from February 2024 (see \Cref{eofproviders}). The current market dominance by three players, who build approximately \SI{80}{\%} of MEV-Boost blocks (see \Cref{bms}), raises the barrier of entry for builders, requiring them to secure order flow deals and operate their own searchers to become competitive \cite{builder_dominance}. Builders unable to do so are forced to adopt subsidy strategies, which are unsustainable in the long run.

Furthermore, builders with a latency advantage due to advanced infrastructure, such as \gls{hft} firms, can scale up by running multiple instances and adjusting bid values until the last milliseconds of the block auction (see \Cref{latency}) . This gives them an edge in reacting to others' bids and placing cancellations, at the cost of further raising barriers to entry, misaligning validators' incentives, and increasing the auction's gameability \cite{cancel}.

To uphold Ethereum's censorship resistance and neutrality, it is essential to foster competition and maintain a sufficiently decentralized builder market. There are multiple solution directions for achieving this:

\begin{itemize}
    \item \textbf{Changing Auction Format}:  To diminish the impact of latency and bidding strategies, the block auction format can be changed to a sealed-bid auction, eliminating adaptive bidding strategies \cite{wu_strategic_2023}. However, such an opaque design requires further consideration to reduce dependency on relay intermediaries for not disclosing bid values while maintaining an efficient auction, secure against collusion between the builders and the relays via side-channels.
    \newpage
    \item \textbf{Decentralized Builders}: Solutions to decentralize the block builder role, such as SUAVE \cite{noauthor_suave_2022},\footnote{SUAVE is currently under development \cite{noauthor_flashbotssuave-specs_2024}.} aim to remove social trust in builder by running block building logic in privacy-preserving execution environments like \glspl{tee}. They can contribute to a less monopolized builder market if they become competitive by attracting a diverse set of order flow providers. Features like gas fee refunds \cite{noauthor_fbgas_2024}, making searchers pay the second-best price of the bundles competing for access to the same state, can also help achieve this.

    \item \textbf{Redistributing and Reducing \gls{mev}}: Capturing the \gls{mev} extracted by searchers and redistributing it could decrease the value leaked to builders and, eventually, to validators. Solutions for this include \gls{mev}-aware decentralized applications (e.g., \glspl{dex} \cite{noauthor_1inch_nodate ,noauthor_cow_nodate, noauthor_sorella_nodate, uniswapX}), intermediary protocols like \glspl{ofa} \cite{mev_blocker, noauthor_mev-share_2024}, and further mechanisms such as \gls{mev} taxes \cite{noauthor_priority_2024}. Additionally, encrypted mempools (e.g., Shutter Network \cite{shutter_shutter_nodate}, SUAVE \cite{noauthor_suave_2022}) where users send encrypted transactions that are decyrpted before execution, can prevent censoring and potentially reduce the \gls{mev} exposed.

    \item \textbf{In-Protocol Mechanisms}: In-protocol mechanisms such as inclusion lists \cite{noauthor_il_2023, noauthor_focil_2024} and concurrent proposers \cite{fox_censorship_2023, noauthor_concurrent_2024} enforce the involvement of a set of transactions/bundles in a produced block. This could strengthen Ethereum's censorship resistance properties by reducing the monopoly of a single builder over the block payload, making block production a cooperative activity. Moreover, further unbundling consensus from execution by clearly separating block proposal and attesting duties, referred to as \gls{aps}, would allow shielding attesters from centralizing forces. Both execution tickets \cite{noauthor_executiontickets_2023} and execution auctions \cite{noauthor_execution_2024} are proposals in that direction, and are discussed along with inclusion lists to prevent timing games \cite{schwarz-schilling_time_2023, oz_time_2023} without introducing new \gls{mev} vectors \cite{noauthor_mevresistant_2024}.

\end{itemize}

Overall, the discussed solutions propose various ways to address the censorship threat posed by the currently centralized block builder market. We hope our study provides valuable insights and considerations for designing further iterations of such mechanisms and Ethereum block auctions, preserving Ethereum's censorship resistance properties.

\section{Conclusion}
In this paper, we examined the MEV-Boost auction to identify features significant for winning blocks and earning profits. We found that builders' order flow diversity and access to order flow by \glspl{ep} are correlated with their block market share and profitability, respectively. Additionally, we showed a positive correlation between market share and profit margin among the top ten builders, suggesting a ``chicken-and-egg'' problem where builders need order flow distinguishing them from others to profit but only receive such flow if they have a significant market share. Finally, we discussed how the key features we identified for success in MEV-Boost raise the barrier of entry for new builders, driving the builder market towards centralization, and explored existing solutions for preserving Ethereum's censorship resistance properties, hoping the insights we provide help move the solution space forward.

\section*{Acknowledgements}
We thank The Latest in DeFi Research (TLDR) program, funded by the Uniswap Foundation, for supporting this work. Special thanks to Flashbots and Eden for generously providing data. We also express our gratitude for the valuable reviews of Julian Ma, Xin Wan, Christof Ferreira Torres, Kevin Pang, Žan Knafelc, and Tesa Ho.

\bibliography{sample-bibliography}

\appendix

\section{Additional Methodology}
\subsection{Label Identification}\label{heuristics}
\subsubsection{OFA Bundles}

We consider an \gls{ofa} bundle structure of a user transaction $t^i$ at block position index $i$, a backrun transaction $t^{i+1}$, and a builder refund transaction $t^{i+2}$ which rebates a proportion of the backrun's paid fee to the user or the \gls{ofa}, which then distributes it. We also identify cases where an \gls{ofa} bundle is backrunned by another \gls{ofa} bundle structure, only involving a backrun transaction and a refund, without a user transaction.\footnote{Example of a two-transaction \gls{ofa} bundle:
\begin{itemize}
    \item Backrun: \texttt{0x2a6b88d73d939310e59b569ef7acf0fe6ae9df97db5d1a2a8006218d1a713cd2}
    \item Refund: \texttt{0x7e5b3622c83694aabbc298df1c524d1b02736a995dd25deb20353f652ed29683}
\end{itemize}} Overall, we identify three bundle structures, examining adjacent transactions by index, with the following heuristics:

\begin{itemize}
    \item \textit{Cowswap-MEVBlocker Bundle}
    \begin{enumerate}
            \item $t$ is issued to the Cowswap settlement contract,\footnote{Cowswap Protocol Settlement address: \texttt{0x9008D19f58AAbD9eD0D60971565AA8510560ab41}}
            \item $t^{i+1}_{priorityfee}+t^{i+1}_{coinbase} \geq t^{i+2}_{value}$,\footnote{This condition is enforced to ensure that the builder's refund transaction transfers at most the complete fee payment it received from the backrun transaction.}
             \item $t^{i+2}$ transfers value to the MEVBlocker Gnosis Safe address,\footnote{MEVBlocker Gnosis Safe address: \texttt{0xce91228789b57deb45e66ca10ff648385fe7093b}}
            \item $t^{i+2}$ is issued by the builder of the block.
    \end{enumerate}
    \item \textit{Matching Address Bundle}
    \begin{enumerate}
            \item $t$ involves an ERC-20 transfer or a swap,
            \item $t^{i+1}_{priorityfee}+t^{i+1}_{coinbase} \geq t^{i+2}_{value}$,
             \item $t^{i+2}$ transfers value to the sender of $t$,
            \item $t^{i+2}$ is issued by the builder of the block.
    \end{enumerate}
    \item \textit{MEV-Share Bundle}
    \begin{enumerate}
            \item $t$ involves an ERC-20 transfer or a swap,
            \item $t^{i+1}_{priorityfee}+t^{i+1}_{coinbase} \geq t^{i+2}_{value}$,
             \item $t^{i+2}$ transfers value to a known MEV-Share refund address \cite{fb_protect_data},
            \item $t^{i+2}$ is issued by the builder of the block.
    \end{enumerate}
\end{itemize}

\subsubsection{MEV}
We detect certain \gls{mev} strategies, such as atomic arbitrages, sandwiches, and liquidations, using the labels provided by zeromev\cite{noauthor_zeromev_nodate}. For arbitrages and sandwiches, we combine zeromev data with labels available on Dune Analytics collections \cite{github_atomic_mev_arb, github_atomic_mev_sandwiches, github_atomic_mev_sandwiched}. We programmatically identify all victim transactions in a sandwich bundle.
\subsubsection{Non-Atomic Arbitrages}
We detect non-atomic \gls{cex}-\gls{dex} arbitrages, introducing heuristics inspired from \cite{colincexdex,searcherbuilder,heimbach_non-atomic_2024}. We label a transaction $t^i$ at a normalized block position $i$ as a non-atomic arbitrage if it fulfills \textit{all} of the following heuristics:
\begin{enumerate}
    \item $t$ has a sufficient gas fee or does coinbase payment \cite{heimbach_non-atomic_2024}:\footnote{If $t^{i-1}$ or $t^{i+1}$ has the same sender as $t$ and make a coinbase transfer, we also consider it as a payment for $t$ \cite{colincexdex, searcherbuilder}}
$t_{tip} \geq 1 Gwei$ or $t_{coinbase} > 0$ ,
    \item $t$ is not seen in the public mempool,
    \item $t$ does not have an \gls{mev} label and is not part of an \gls{ofa} bundle,
    \item $t$ makes strictly two ERC-20 token transfers (or one swap) and uses less than \SI{400000}{} gas \cite{heimbach_non-atomic_2024},
    \item $t$ is the first swap in the pool considering trade direction (i.e., the first sell and buy are both seen as the first swap) \cite{heimbach_non-atomic_2024},
    \item $t$ is not submitted to a known router smart contract identified in \cite{methodology_orderflow}.
\end{enumerate}
\subsubsection{Telegram Bot Flow}
We use a dataset available on Dune Analytics \cite{query_telegram_bot} to identify the smart contracts and trades of \SI{21}{} Telegram bots and one web trading portal.
\subsubsection{Solver Model Flow}
In this work, we focus on Cowswap\cite{noauthor_cow_nodate}, UniswapX\cite{uniswapX}, and 1inch Fusion\cite{noauthor_1inch_nodate} solver models. Since solvers and fillers operating on them submit their solutions to the router contracts of these platforms, we detect solver model flow by checking the destination address of the transactions. We identify such router contracts using the Dune Analytics dataset by \cite{methodology_orderflow}.
\subsubsection{CEX}
We detect \gls{cex} deposits utilizing a collection available on Dune Analytics \cite{query_cex_label}. This dataset includes \SI{1922}{} deposit wallet addresses of \SI{301}{} \glspl{cex}.
\subsubsection{Swap Flow}
We identify transactions that involve an ERC-20 transfer (or a swap) and do not belong to other order flow label categories as swap flow. We differentiate between retail and bot swap by checking whether the transaction is directed to a known, non-\gls{mev} smart contract, indicating a retail swap. For this, we use the router addresses made available by \cite{methodology_orderflow} and labeled contracts dataset \cite{noauthor_searcherbuilderpicslabelsnon_mev_contractspy_nodate} used in \cite{searcherbuilder}. The retail swap category also involves the bundled user transactions in \glspl{ofa} and sandwiches. If the transaction does not initially interact with a labeled contract, we categorize it as bot swap.
\subsubsection{Others}
We label the transactions that do not belong to any other order flow category as other public or exclusive flow based on their visibility in the network, using the mempool data by \cite{github_mempool_dumpster}.

\subsection{External Data}\label{externaldata}

As part of our methodology, we used datasets from external resources, summarized in \Cref{tab:datasets}.

\section{Further Measurements}\label{all_builders_appendix}

\subsection{Order Flow Overview}\label{of_appendix}
In \Cref{tab:of_stats}, we summarize our measurements for transparency and order flow labels. 
\subsubsection{Occurance and Trends}
We find that Telegram bot transactions appear more often in blocks (\SI{0.89}{} per block) than common \gls{mev} strategies such as sandwiches (\SI{0.58}{}) and arbitrages (atomic: \SI{0.16}{}, non-atomic: \SI{0.68}{}), as summarized in \Cref{tab:of_stats}. The frequency of occurrences we calculate for \gls{mev} order flow types differs from \cite{heimbach_non-atomic_2024}, which covers MEV-Boost blocks up to October 31, 2023. As we show in \Cref{fig:of_prominence}, the significance of order flow labels changes over time, with certain types becoming more popular. For example, Telegram bot transactions consistently provided the most value until the last week of 2023. Afterwards, we observe oscillations, with sandwiches and non-atomic arbitrages gaining prominence. These trends in the order flow landscape could explain the different measurements compared to \cite{heimbach_non-atomic_2024}.

Interestingly, the gas consumed by other public flow decreases in March 2024 (see \Cref{fig:of_prominence}). As discussed in \Cref{ofl}, this order flow includes transactions such as batch submissions by rollup solutions, potentially consuming high amounts of gas. Following the Dencun upgrade on March 13, 2024 \cite{dencun}, which introduced blob transactions to save gas costs for rollups, we notice a drop in the daily average gas used by other public flow from \SI{35}{\%} to \SI{27.5}{\%}. This is another evidence of how the order flow landscape can evolve over time.

\subsection{All Builders Overview}\label{all_overview}
In \Cref{tab:all_stats}, we summarize all builders' market share, profitability, and block packing metrics.

\subsection{Builder Order Flow}\label{detailed_of_breakdown}
\begin{figure}[t!]
    \centering
        \includegraphics[width=0.7\linewidth]{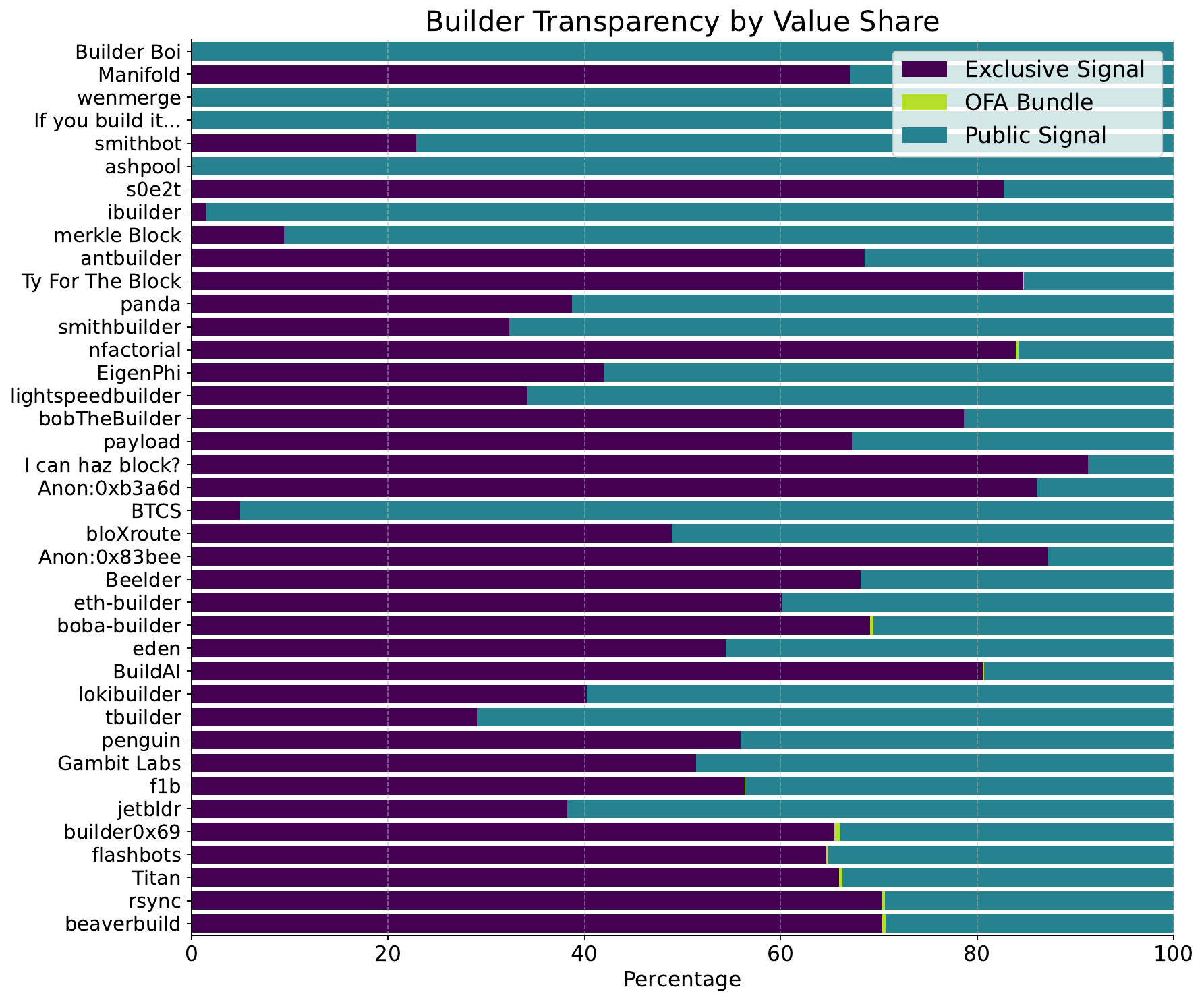}
        \caption{A horizontal bar plot indicating the share of value builders receive from each transparency label. Builders on the y-axis are ordered in ascending order based on their market share, with the builder with the highest share listed at the bottom.}\label{fig:all_builder_transparancy}
\end{figure}

\begin{figure}[t!]
    \centering
    \includegraphics[width=0.7\linewidth]{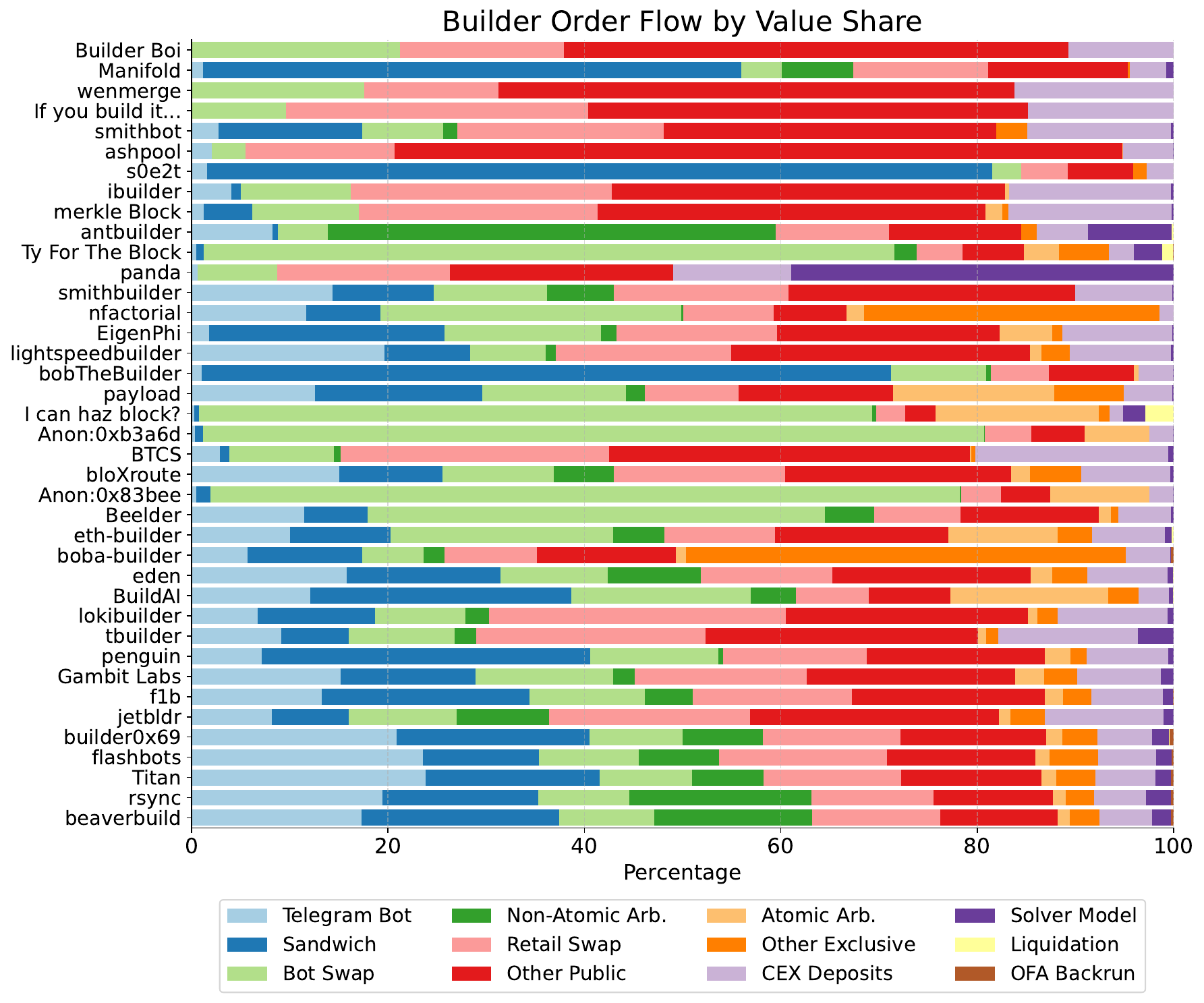}
    \caption{A horizontal bar plot showing the share of value builders receive from each order flow label. Builders on the y-axis are ordered in ascending order based on their market share, with the builder with the highest share listed at the bottom.}\label{fig:all_builder_of}
\end{figure}

In \Cref{fig:all_builder_transparancy} and \Cref{fig:all_builder_of}, we show the order flow transparency and composition of every builder, respectively.

\subsection{Exclusive Order Flow Providers} \label{all_eof}
\begin{figure}[t!]
    \centering
    \begin{subfigure}[b]{0.35\linewidth}
        \includegraphics[width=\linewidth]{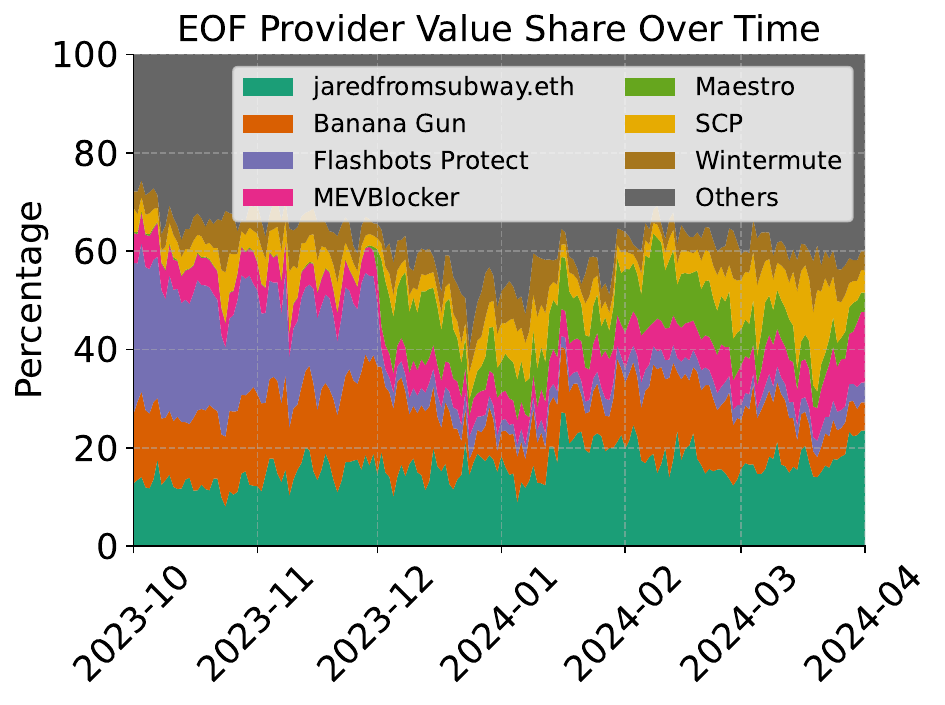}
         \caption{}\label{fig:eof_tseries}
    \end{subfigure}
     \begin{subfigure}[b]{0.35\linewidth}
        \includegraphics[width=\linewidth]{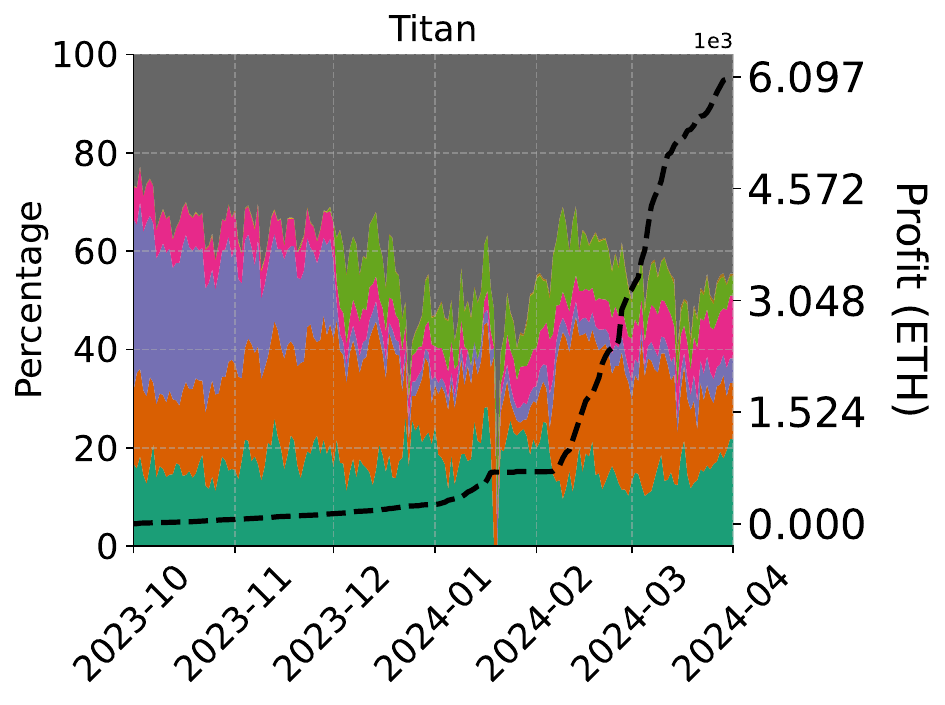}
     \caption{}\label{fig:titan}
    \end{subfigure}
    \vspace{10pt}
      \begin{subfigure}[b]{0.35\linewidth}
        \includegraphics[width=\linewidth]{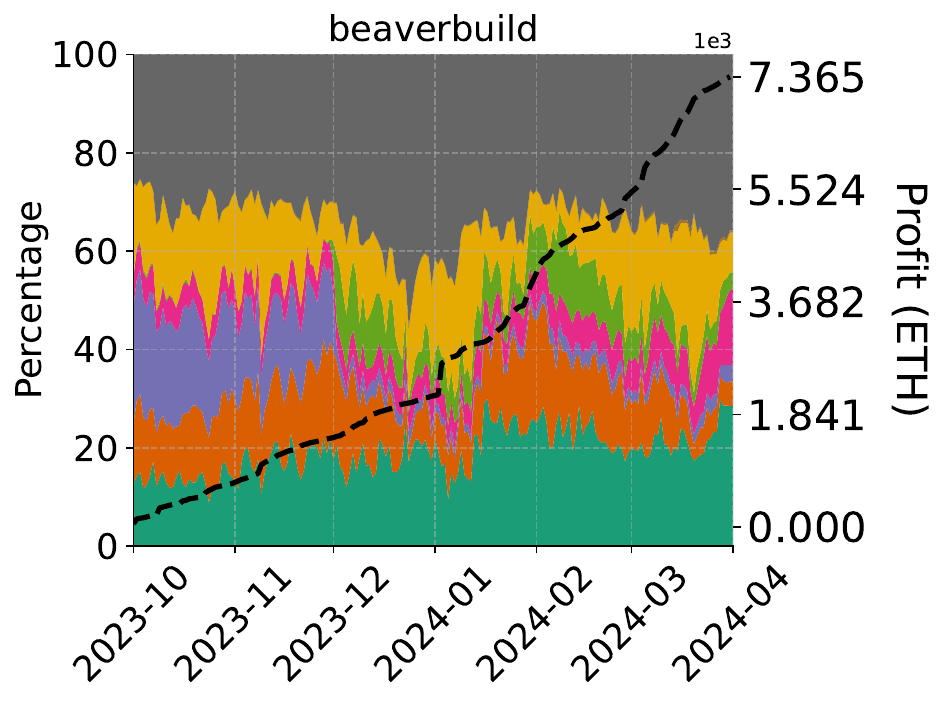}
     \caption{}\label{fig:beaver}
    \end{subfigure}
      \begin{subfigure}[b]{0.35\linewidth}
        \includegraphics[width=\linewidth]{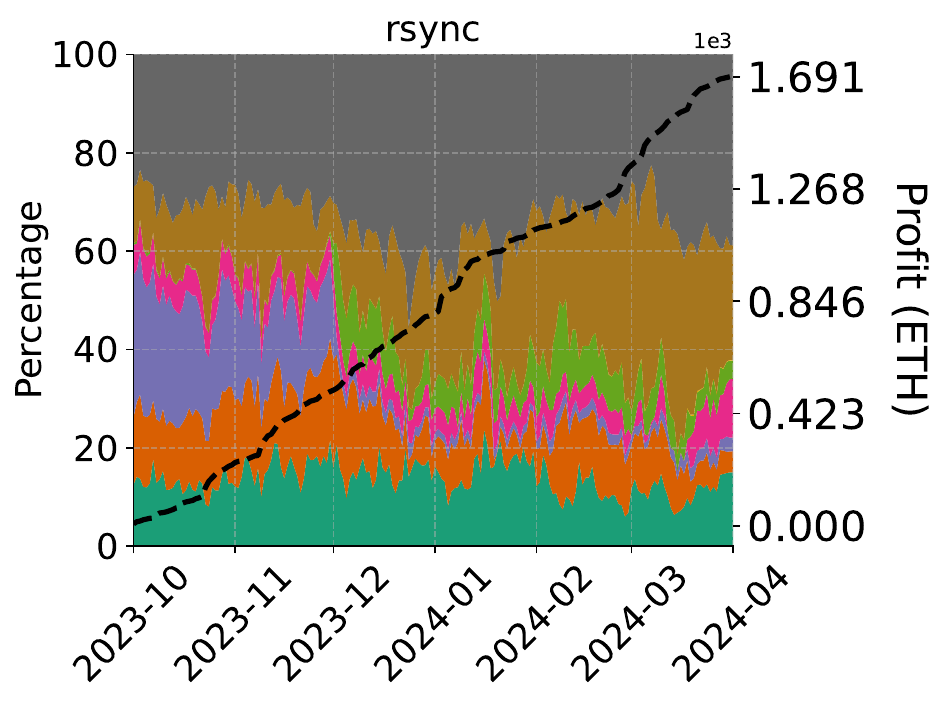}
     \caption{}\label{fig:rsync}
    \end{subfigure}
     \caption{The significance of \gls{eof} providers. \Cref{fig:eof_tseries} is an area plot showing the trends in the share of value contributed by the seven most prominent \gls{eof} providers based on total value. The total \gls{eof} value of the remaining providers is aggregated and denoted as ``Others'' in the plot legend. \Cref{fig:titan}, \Cref{fig:beaver}, and \Cref{fig:rsync} are \gls{eof} provider area plots for \texttt{Titan}, \texttt{beaverbuild}, and \texttt{rysnc} builders, respectively, with a dashed line plot showing their cumulative profit in ETH on the right y-axis. The legend in \Cref{fig:eof_tseries} applies to all figures.
     }\label{fig:eof}
\end{figure}

In \Cref{fig:eof_tseries}, we show the trends in the significance of the seven \gls{eof} providers who contributed the most value to MEV-Boost blocks. The emergence of the Maestro Telegram bot coincides with a major decrease in the significance of Flashbots Protect and MEVBlocker, supporting the findings in \cite{yang_decentralization_2024}. Additionally, Banana Gun Telegram bot gains a traction at the beginning of February 2024, which we observe to be overlapping with the surge in \texttt{Titan}'s market share (see \Cref{fig:market_share}) and profits (see \Cref{fig:titan}), suggesting an \gls{eof} deal.\footnote{The spike in \Cref{fig:titan}, observed on January 19, 2024, is due to a failing \gls{rpc} endpoint of \texttt{Titan} \cite{titan_builder__titanbuilderxyz_we_2024}.} Conversely, \texttt{beaverbuild} and \texttt{rsync} consistently made profits over time, likely due to their integrated searchers SCP and Wintermute \cite{gupta_centralizing_2023, heimbach_non-atomic_2024}, respectively (see \Cref{fig:beaver}, \Cref{fig:rsync}).

\begin{figure}[t!]
\centering
\includegraphics[width=0.7\linewidth]{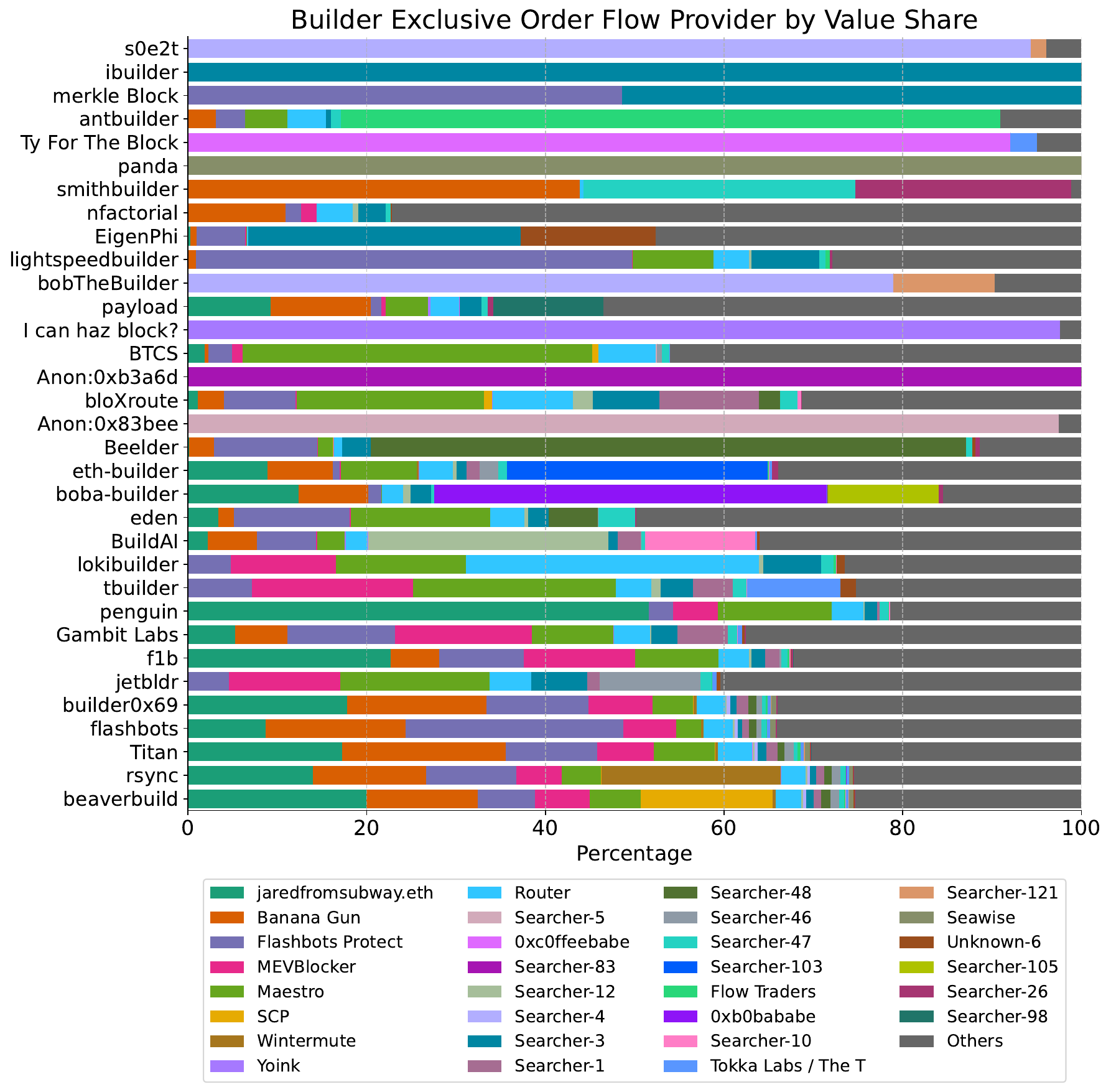}
\caption{A horizontal bar plot showing the share \gls{eof} value builders receive from providers. Builders on the y-axis are ordered in ascending order based on their market share, with the builder with the highest share listed at the bottom. Unknown providers are enumerated according to \cite{eof_addresses}.}\label{fig:all_eof_value_ratio}
\end{figure}

In \Cref{fig:all_eof_value_ratio}, we present the significance of \gls{eof} providers for MEV-Boost builders with greater than \SI{0.01}{\%} market share. As we could not match most of the providers with known entities, we enumerate them and share this mapping in \cite{eof_addresses}.

\begin{table}[t!]
\centering
\footnotesize
\setlength{\tabcolsep}{2.5pt}
\begin{tabular}{@{}lr@{}} 
\toprule
Data & Source \\ \midrule
Certain MEV and swap transactions & zeromev \cite{noauthor_zeromev_nodate} \\
Non-MEV smart contract addresses & searcherbuilder.pics \cite{searcherbuilder} \\
Searcher addresses & Etherscan \cite{etherscanio_ethereum_nodate}, Arkham \cite{noauthor_arkham_nodate}, libMEV \cite{noauthor_libmev_nodate} \\
Bids submitted to the UltraSound relay  & Eden Network \cite{edendocs} \\
Payloads of all MEV-Boost auctions & relayscan.io \cite{relayscan} \\
Public keys controlled by the builders & Builder Websites, Eden Network \cite{edendocs} \\ 
MEVBlocker transactions/bundles & Dune Analytics \cite{noauthor_mevblockerdune_2024} \\
Flashbots Protect and MEV-Share transactions/bundles & Flashbots Data \cite{fb_protect_data} \\
\bottomrule
\end{tabular}
\caption{External Data Sources}\label{tab:datasets}
\end{table}


\subsection{Latency and Bidding}\label{all_latency}
In \Cref{tab:all_latency}, we summarize the latency and bidding behavior of all builders for the blocks they won on the UltraSound relay.

\subsection{Validator Payment}\label{validator_payment_appendix}
In \Cref{tab:validator_payment_patterns}, we summarize builders' validator payment patterns. Excluded blocks correspond to MEV-Boost blocks where the builder sets the proposer as the fee recipient. Other fee payer blocks are the ones where the fee recipient builder address does not conduct the proposer payment, but a related address does it. We include such blocks in our analysis as the builder is assumed to eventually pay the original payer. Further, we measure the difference between the value indicated in the blocks' payload value obtained from MEV-Boost relays and the actual value paid to the validator. Here, we calculate the over-promised value (promised but not paid to the proposer) and the under-promised value (paid to the proposer on top of the promised value). Finally, we calculate the relay payment done by the builder, which occurs if the builder uses UltraSound's bid adjustment feature \cite{ultrasound_bid}.

We observe that \texttt{Titan} produced many blocks which were excluded from our analysis. While \texttt{Titan} did not receive on-chain value from these blocks, they helped the builder to gain market share, as discussed in \Cref{validator_payment}. Moreover, \texttt{flashbots} and \texttt{penguin} have over-promised a considerable amount, potentially by including withdrawal transactions in their blocks. Finally, we detect that only \gls{brt} used the bid adjustment feature of the UltraSound relay, as no other builders made a relay payment.

\begin{table}[t!]
\centering
\footnotesize
\begin{tabular}{clccccr}
\toprule
& & Occurrence & Avg. & Avg.  & Value & Avg. \\ 
Type & Label & Probability & Value  & Share & per Gas & Gas \\
& &  $P$ &  [ETH]  & [\%] &  [ETH] & [\%] \\ 
\midrule
& Exclusive Signal  & 0.99 & 0.12 & 66.69 & $7.5e-09$ & 19.6 \\ 
Transparency & Public Signal  & 1.0 & 0.02 & 33.0 & $1.3e-09$ & 80.26 \\ 
& OFA Bundle  & 0.04 & $<0.01$ & 0.31 & $1.8e-11$ & 0.14 \\
\cdashlinelr{1-7}
& Telegram Bot Flow  & 0.89 & 0.04 & 18.64 & $2.4e-09$ & 5.93 \\ 
& Sandwich MEV  & 0.58 & 0.03 & 17.49 & $1.8e-09$ & 1.7 \\ 
& Bot Swap Flow  & 1.0 & 0.02 & 10.09 & $1.2e-09$ & 12.69 \\ 
& Non-Atomic Arbitrage MEV  & 0.68 & 0.02 & 13.22 & $1.2e-09$ & 1.52 \\ 
& Retail Swap Flow  & 1.0 & 0.01 & 13.89 & $5.9e-10$ & 33.33 \\ 
Order & Other Public Flow & 1.0 & 0.01 & 13.73 & $5.6e-10$ & 34.28 \\ 
Flow & Atomic Arbitrage MEV   & 0.16 & 0.01 & 1.49 & $3.4e-10$ & 0.41 \\ 
& Other Exclusive Flow  & 0.87 & 0.01 & 3.34 & $3.4e-10$ & 1.44 \\ 
& CEX Deposits  & 0.99 & $<0.01$ & 5.96 & $2e-10$ & 7.22 \\ 
& Solver Model Flow  & 0.52 & $<0.01$ & 1.9 & $1.1e-10$ & 1.42 \\ 
& Liquidation MEV  & $<0.01$ & $<0.01$ & 0.03 & $6.7e-11$ & 0.01 \\ 
& OFA Backrun  & 0.04 & $<0.01$ & 0.22 & $1.4e-11$ & 0.05 \\ 
\bottomrule
\end{tabular}
\caption{Summary of Transparency and Order Flow Labels} 
\label{tab:of_stats}
\end{table}

\begin{landscape}
\begin{table}[htbp]
\centering
\footnotesize
\setlength{\tabcolsep}{2.5pt}
\begin{tabular}{lcccccccccccc}
\toprule
 & Total  & Market & Total & Total & Total & Profit & Profitable & Dust/Zero Profit & Subsidized & ToB & BoB & EoB \\
Builder & Blocks & Share & Payment & Subsidy & Profit & Margin & Blocks & Blocks & Blocks & Value & Value & Value \\
\ & [\#] & [\%] & [ETH] & [ETH] & [ETH] & [\%] & [\%] & [\%] & [\%] & [\%] & [\%] & [\%] \\ 
\midrule
beaverbuild & \SI{413868}{} & \SI{37.91}{} & \SI{49871.82}{} & \SI{-70.46}{} & \SI{7341.62}{} & \SI{5.4}{} & \SI{45.11}{} & \SI{46.26}{} & \SI{8.63}{} & \SI{74.55}{} & \SI{25.17}{} & \SI{0.27}{} \\ 
rsync & \SI{273126}{} & \SI{25.02}{} & \SI{33691.74}{} & \SI{-80.71}{} & \SI{1679.73}{} & \SI{3.25}{} & \SI{39.05}{} & \SI{58.6}{} & \SI{2.35}{} & \SI{74.15}{} & \SI{25.18}{} & \SI{0.66}{} \\ 
Titan & \SI{177915}{} & \SI{16.3}{} & \SI{22025.19}{} & \SI{-61.49}{} & \SI{6083.95}{} & \SI{1.02}{} & \SI{27.42}{} & \SI{52.48}{} & \SI{20.1}{} & \SI{74.91}{} & \SI{20.5}{} & \SI{4.59}{} \\ 
flashbots & \SI{74636}{} & \SI{6.84}{} & \SI{6918.79}{} & \SI{0.0}{} & \SI{48.22}{} & \SI{1.38}{} & \SI{15.66}{} & \SI{84.34}{} & \SI{0.0}{} & \SI{72.32}{} & \SI{27.31}{} & \SI{0.37}{} \\ 
builder0x69 & \SI{35083}{} & \SI{3.21}{} & \SI{3878.92}{} & \SI{0.0}{} & \SI{434.64}{} & \SI{2.3}{} & \SI{23.75}{} & \SI{76.25}{} & \SI{0.0}{} & \SI{71.06}{} & \SI{28.65}{} & \SI{0.28}{} \\ 
jetbldr & \SI{32670}{} & \SI{2.99}{} & \SI{1192.15}{} & \SI{-73.21}{} & \SI{11.41}{} & \SI{-12.45}{} & \SI{6.21}{} & \SI{12.67}{} & \SI{81.13}{} & \SI{66.96}{} & \SI{32.64}{} & \SI{0.4}{} \\ 
f1b & \SI{16716}{} & \SI{1.53}{} & \SI{1251.63}{} & \SI{-12.39}{} & \SI{53.59}{} & \SI{-1.94}{} & \SI{16.72}{} & \SI{42.01}{} & \SI{41.27}{} & \SI{65.48}{} & \SI{34.03}{} & \SI{0.49}{} \\ 
Gambit Labs & \SI{15281}{} & \SI{1.4}{} & \SI{889.45}{} & \SI{-19.75}{} & \SI{78.42}{} & \SI{-4.03}{} & \SI{21.26}{} & \SI{24.72}{} & \SI{54.02}{} & \SI{57.41}{} & \SI{34.93}{} & \SI{7.66}{} \\ 
penguin & \SI{14622}{} & \SI{1.34}{} & \SI{1028.28}{} & \SI{-41.77}{} & \SI{17.7}{} & \SI{-9.73}{} & \SI{11.8}{} & \SI{10.26}{} & \SI{77.94}{} & \SI{29.61}{} & \SI{66.7}{} & \SI{3.7}{} \\ 
tbuilder & \SI{10584}{} & \SI{0.97}{} & \SI{338.38}{} & \SI{-79.1}{} & \SI{-77.01}{} & \SI{-41.23}{} & \SI{0.44}{} & \SI{1.67}{} & \SI{97.88}{} & \SI{63.66}{} & \SI{36.07}{} & \SI{0.27}{} \\ 
lokibuilder & \SI{5609}{} & \SI{0.51}{} & \SI{253.18}{} & \SI{-10.52}{} & \SI{36.21}{} & \SI{-2.86}{} & \SI{23.28}{} & \SI{16.51}{} & \SI{60.21}{} & \SI{66.81}{} & \SI{32.13}{} & \SI{1.06}{} \\ 
BuildAI & \SI{3657}{} & \SI{0.33}{} & \SI{1054.98}{} & \SI{0.0}{} & \SI{196.48}{} & \SI{13.49}{} & \SI{74.84}{} & \SI{25.16}{} & \SI{0.0}{} & \SI{83.31}{} & \SI{16.53}{} & \SI{0.16}{} \\ 
eden & \SI{2289}{} & \SI{0.21}{} & \SI{185.06}{} & \SI{0.0}{} & \SI{1.55}{} & \SI{2.0}{} & \SI{18.17}{} & \SI{81.83}{} & \SI{0.0}{} & \SI{66.08}{} & \SI{32.54}{} & \SI{1.38}{} \\ 
boba-builder & \SI{2126}{} & \SI{0.19}{} & \SI{412.33}{} & \SI{-2.6}{} & \SI{1.12}{} & \SI{-3.29}{} & \SI{26.76}{} & \SI{44.59}{} & \SI{28.65}{} & \SI{34.21}{} & \SI{21.25}{} & \SI{44.54}{} \\ 
eth-builder & \SI{1922}{} & \SI{0.18}{} & \SI{539.45}{} & \SI{-2.24}{} & \SI{159.71}{} & \SI{-11.67}{} & \SI{40.17}{} & \SI{31.53}{} & \SI{28.3}{} & \SI{59.87}{} & \SI{35.47}{} & \SI{4.66}{} \\ 
Beelder & \SI{1751}{} & \SI{0.16}{} & \SI{155.42}{} & \SI{0.0}{} & \SI{1.11}{} & \SI{1.2}{} & \SI{16.62}{} & \SI{83.38}{} & \SI{0.0}{} & \SI{79.24}{} & \SI{20.26}{} & \SI{0.5}{} \\ 
Anon:0x83bee & \SI{1631}{} & \SI{0.15}{} & \SI{1519.58}{} & \SI{0.0}{} & \SI{427.16}{} & \SI{23.49}{} & \SI{100.0}{} & \SI{0.0}{} & \SI{0.0}{} & \SI{36.19}{} & \SI{54.33}{} & \SI{9.48}{} \\ 
bloXroute & \SI{1134}{} & \SI{0.1}{} & \SI{120.22}{} & \SI{-1.37}{} & \SI{1.84}{} & \SI{-6.42}{} & \SI{23.37}{} & \SI{35.19}{} & \SI{41.45}{} & \SI{68.87}{} & \SI{24.35}{} & \SI{6.78}{} \\ 
Anon:0xb3a6d & \SI{997}{} & \SI{0.09}{} & \SI{974.33}{} & \SI{0.0}{} & \SI{215.69}{} & \SI{25.93}{} & \SI{98.4}{} & \SI{1.6}{} & \SI{0.0}{} & \SI{49.89}{} & \SI{45.92}{} & \SI{4.19}{} \\ 
BTCS & \SI{955}{} & \SI{0.09}{} & \SI{19.59}{} & \SI{-8.97}{} & \SI{-8.9}{} & \SI{-226.95}{} & \SI{1.78}{} & \SI{5.55}{} & \SI{92.67}{} & \SI{53.76}{} & \SI{45.86}{} & \SI{0.38}{} \\ 
I can haz block? & \SI{527}{} & \SI{0.05}{} & \SI{1656.91}{} & \SI{-1.23}{} & \SI{1094.65}{} & \SI{40.26}{} & \SI{97.34}{} & \SI{1.9}{} & \SI{0.76}{} & \SI{64.17}{} & \SI{35.54}{} & \SI{0.29}{} \\ 
payload & \SI{426}{} & \SI{0.04}{} & \SI{312.69}{} & \SI{0.0}{} & \SI{14.23}{} & \SI{11.18}{} & \SI{56.1}{} & \SI{43.9}{} & \SI{0.0}{} & \SI{78.11}{} & \SI{20.23}{} & \SI{1.65}{} \\ 
bobTheBuilder & \SI{408}{} & \SI{0.04}{} & \SI{130.09}{} & \SI{0.0}{} & \SI{49.11}{} & \SI{9.16}{} & \SI{44.12}{} & \SI{55.88}{} & \SI{0.0}{} & \SI{77.65}{} & \SI{21.69}{} & \SI{0.65}{} \\ 
lightspeedbuilder & \SI{355}{} & \SI{0.03}{} & \SI{21.49}{} & \SI{-0.86}{} & \SI{-0.76}{} & \SI{-13.29}{} & \SI{1.41}{} & \SI{6.76}{} & \SI{91.83}{} & \SI{67.97}{} & \SI{30.86}{} & \SI{1.17}{} \\ 
EigenPhi & \SI{343}{} & \SI{0.03}{} & \SI{19.75}{} & \SI{-0.0}{} & \SI{0.28}{} & \SI{3.07}{} & \SI{23.91}{} & \SI{75.8}{} & \SI{0.29}{} & \SI{66.77}{} & \SI{31.43}{} & \SI{1.8}{} \\ 
nfactorial & \SI{335}{} & \SI{0.03}{} & \SI{112.33}{} & \SI{0.0}{} & \SI{63.27}{} & \SI{29.43}{} & \SI{97.31}{} & \SI{2.69}{} & \SI{0.0}{} & \SI{10.44}{} & \SI{83.24}{} & \SI{6.32}{} \\ 
smithbuilder & \SI{330}{} & \SI{0.03}{} & \SI{22.12}{} & \SI{-0.64}{} & \SI{-0.65}{} & \SI{-12.34}{} & \SI{0.0}{} & \SI{10.61}{} & \SI{89.39}{} & \SI{53.23}{} & \SI{39.33}{} & \SI{7.43}{} \\ 
panda & \SI{286}{} & \SI{0.03}{} & \SI{23.74}{} & \SI{-15.7}{} & \SI{-15.69}{} & \SI{-246.71}{} & \SI{0.35}{} & \SI{1.4}{} & \SI{98.25}{} & \SI{72.04}{} & \SI{27.5}{} & \SI{0.46}{} \\ 
Ty For The Block & \SI{276}{} & \SI{0.03}{} & \SI{495.89}{} & \SI{-0.06}{} & \SI{788.55}{} & \SI{40.04}{} & \SI{94.2}{} & \SI{3.62}{} & \SI{2.17}{} & \SI{79.65}{} & \SI{19.5}{} & \SI{0.85}{} \\ 
antbuilder & \SI{207}{} & \SI{0.02}{} & \SI{89.94}{} & \SI{-0.43}{} & \SI{2.27}{} & \SI{6.26}{} & \SI{64.73}{} & \SI{21.26}{} & \SI{14.01}{} & \SI{85.73}{} & \SI{14.04}{} & \SI{0.23}{} \\ 
merkle Block & \SI{163}{} & \SI{0.01}{} & \SI{3.7}{} & \SI{-0.8}{} & \SI{-0.75}{} & \SI{-49.37}{} & \SI{0.61}{} & \SI{0.61}{} & \SI{98.77}{} & \SI{63.22}{} & \SI{35.99}{} & \SI{0.8}{} \\ 
ibuilder & \SI{149}{} & \SI{0.01}{} & \SI{2.6}{} & \SI{-0.44}{} & \SI{-0.44}{} & \SI{-24.63}{} & \SI{0.0}{} & \SI{0.67}{} & \SI{99.33}{} & \SI{50.31}{} & \SI{49.17}{} & \SI{0.52}{} \\ 
s0e2t & \SI{122}{} & \SI{0.01}{} & \SI{24.4}{} & \SI{0.0}{} & \SI{0.11}{} & \SI{1.43}{} & \SI{15.57}{} & \SI{84.43}{} & \SI{0.0}{} & \SI{22.79}{} & \SI{76.3}{} & \SI{0.91}{} \\ 
ashpool & \SI{43}{} & \SI{0.0}{} & \SI{10.5}{} & \SI{0.0}{} & \SI{0.03}{} & \SI{1.02}{} & \SI{11.63}{} & \SI{88.37}{} & \SI{0.0}{} & \SI{59.24}{} & \SI{20.48}{} & \SI{20.28}{} \\ 
smithbot & \SI{39}{} & \SI{0.0}{} & \SI{26.68}{} & \SI{0.0}{} & \SI{0.03}{} & \SI{-0.38}{} & \SI{28.21}{} & \SI{71.79}{} & \SI{0.0}{} & \SI{57.5}{} & \SI{40.95}{} & \SI{1.55}{} \\ 
If you build it, they will come. & \SI{11}{} & \SI{0.0}{} & \SI{0.32}{} & \SI{0.0}{} & \SI{0.01}{} & \SI{4.73}{} & \SI{9.09}{} & \SI{90.91}{} & \SI{0.0}{} & \SI{43.72}{} & \SI{54.63}{} & \SI{1.65}{} \\ 
wenmerge & \SI{3}{} & \SI{0.0}{} & \SI{0.28}{} & \SI{0.0}{} & \SI{0.0}{} & \SI{1.61}{} & \SI{0.0}{} & \SI{100.0}{} & \SI{0.0}{} & \SI{78.31}{} & \SI{21.63}{} & \SI{0.06}{} \\ 
Manifold & \SI{2}{} & \SI{0.0}{} & \SI{0.15}{} & \SI{0.0}{} & \SI{0.09}{} & \SI{24.04}{} & \SI{50.0}{} & \SI{50.0}{} & \SI{0.0}{} & \SI{80.73}{} & \SI{19.18}{} & \SI{0.1}{} \\ 
Builder Boi & \SI{2}{} & \SI{0.0}{} & \SI{0.04}{} & \SI{0.0}{} & \SI{-0.0}{} & \SI{0.0}{} & \SI{0.0}{} & \SI{100.0}{} & \SI{0.0}{} & \SI{83.47}{} & \SI{16.49}{} & \SI{0.04}{} \\ 
\bottomrule
\end{tabular}
\caption{Market Share, Profitability, and Block Packing Metrics of All Builders}\label{tab:all_stats}
\end{table}
\end{landscape}

\begin{table}[t!]
\centering
\footnotesize
\begin{tabular}{lcccccr}
\toprule
\ & Total & Avg. & Avg. & Avg. & Total & Avg. \\ 
Builder & Blocks  & Bids  & Update Lag  & Winner Time  & Cancels & Cancels  \\ 
\ & [\#] & [\#] & [ms] & [ms] & [\#] & [\#] \\ 
\midrule
beaverbuild & \SI{137721}{} & \SI{31.54}{} & \SI{130.41}{} & \SI{586.59}{} & \SI{248195}{} & \SI{1.74}{} \\ 
rsync & \SI{95310}{} & \SI{25.27}{} & \SI{166.21}{} & \SI{531.38}{} & \SI{212504}{} & \SI{2.29}{} \\ 
Titan & \SI{47539}{} & \SI{31.11}{} & \SI{186.25}{} & \SI{605.83}{} & \SI{91337}{} & \SI{1.78}{} \\ 
flashbots & \SI{32612}{} & \SI{16.25}{} & \SI{411.36}{} & \SI{348.45}{} & \SI{3014}{} & \SI{0.1}{} \\ 
builder0x69 & \SI{23294}{} & \SI{20.64}{} & \SI{183.08}{} & \SI{575.09}{} & \SI{17445}{} & \SI{0.65}{} \\ 
jetbldr & \SI{12960}{} & \SI{35.2}{} & \SI{103.08}{} & \SI{605.75}{} & \SI{904}{} & \SI{0.09}{} \\ 
Gambit Labs & \SI{9530}{} & \SI{15.85}{} & \SI{349.05}{} & \SI{496.3}{} & \SI{492}{} & \SI{0.1}{} \\ 
f1b & \SI{9242}{} & \SI{25.42}{} & \SI{214.18}{} & \SI{680.45}{} & \SI{10494}{} & \SI{0.98}{} \\ 
penguin & \SI{7783}{} & \SI{49.64}{} & \SI{69.92}{} & \SI{652.91}{} & \SI{966}{} & \SI{0.13}{} \\ 
tbuilder & \SI{4786}{} & \SI{6.1}{} & \SI{476.68}{} & \SI{104.71}{} & \SI{81}{} & \SI{0.01}{} \\ 
lokibuilder & \SI{3445}{} & \SI{23.01}{} & \SI{200.14}{} & \SI{398.06}{} & \SI{1176}{} & \SI{0.32}{} \\ 
BuildAI & \SI{1580}{} & \SI{7.39}{} & \SI{220.17}{} & \SI{511.55}{} & \SI{49}{} & \SI{0.03}{} \\ 
boba-builder & \SI{1048}{} & \SI{15.94}{} & \SI{531.17}{} & \SI{-530.01}{} & \SI{57}{} & \SI{0.04}{} \\ 
Beelder & \SI{905}{} & \SI{1.96}{} & \SI{32.19}{} & \SI{383.65}{} & \SI{18}{} & \SI{0.03}{} \\ 
Anon:0x83bee & \SI{753}{} & \SI{9.65}{} & \SI{462.65}{} & \SI{154.63}{} & \SI{34}{} & \SI{0.04}{} \\ 
eth-builder & \SI{613}{} & \SI{4.81}{} & \SI{598.22}{} & \SI{454.77}{} & \SI{9}{} & \SI{0.02}{} \\ 
Anon:0xb3a6d & \SI{452}{} & \SI{9.29}{} & \SI{645.09}{} & \SI{345.08}{} & \SI{93}{} & \SI{0.17}{} \\ 
I can haz block? & \SI{317}{} & \SI{6.46}{} & \SI{1007.86}{} & \SI{-828.63}{} & \SI{27}{} & \SI{0.04}{} \\ 
lightspeedbuilder & \SI{269}{} & \SI{39.82}{} & \SI{70.68}{} & \SI{502.95}{} & \SI{1472}{} & \SI{5.41}{} \\ 
eden & \SI{262}{} & \SI{5.37}{} & \SI{486.39}{} & \SI{217.26}{} & \SI{227}{} & \SI{0.93}{} \\ 
payload & \SI{210}{} & \SI{12.06}{} & \SI{482.52}{} & \SI{204.68}{} & \SI{62}{} & \SI{0.31}{} \\ 
bloXroute & \SI{189}{} & \SI{10.8}{} & \SI{233.64}{} & \SI{342.44}{} & \SI{452}{} & \SI{0.68}{} \\ 
smithbuilder & \SI{174}{} & \SI{27.29}{} & \SI{152.64}{} & \SI{529.19}{} & \SI{1646}{} & \SI{6.12}{} \\ 
Ty For The Block & \SI{173}{} & \SI{11.38}{} & \SI{579.5}{} & \SI{-291.24}{} & \SI{24}{} & \SI{0.15}{} \\ 
nfactorial & \SI{163}{} & \SI{4.0}{} & \SI{814.76}{} & \SI{-527.05}{} & \SI{1}{} & \SI{0.0}{} \\ 
bobTheBuilder & \SI{139}{} & \SI{3.01}{} & \SI{191.11}{} & \SI{733.85}{} & \SI{4}{} & \SI{0.03}{} \\ 
merkle Block & \SI{127}{} & \SI{46.9}{} & \SI{114.19}{} & \SI{43.7}{} & \SI{331}{} & \SI{1.46}{} \\ 
antbuilder & \SI{125}{} & \SI{59.51}{} & \SI{195.21}{} & \SI{588.75}{} & \SI{1682}{} & \SI{10.21}{} \\ 
ibuilder & \SI{48}{} & \SI{9.52}{} & \SI{499.84}{} & \SI{188.19}{} & \SI{3}{} & \SI{0.06}{} \\ 
ashpool & \SI{26}{} & \SI{9.48}{} & \SI{231.43}{} & \SI{19.09}{} & \SI{3}{} & \SI{0.19}{} \\ 
s0e2t & \SI{20}{} & \SI{1.6}{} & \SI{0.0}{} & \SI{150.5}{} & \SI{1}{} & \SI{0.02}{} \\ 
BTCS & \SI{15}{} & \SI{22.98}{} & \SI{118.91}{} & \SI{-507.28}{} & \SI{40}{} & \SI{2.86}{} \\ 
EigenPhi & \SI{8}{} & \SI{21.14}{} & \SI{654.98}{} & \SI{-1367.5}{} & \SI{7}{} & \SI{1.16}{} \\ 
smithbot & \SI{7}{} & \SI{10.83}{} & \SI{821.82}{} & \SI{470.75}{} & \SI{7}{} & \SI{0.58}{} \\ 
Manifold & \SI{2}{} & \SI{41.0}{} & \SI{363.28}{} & \SI{568.5}{} & \SI{2}{} & \SI{1.0}{} \\ 
If you build it, they will come. & \SI{1}{} & \SI{2.0}{} & \SI{431.0}{} & \SI{-3981.0}{} & \SI{0}{} & \SI{0.0}{} \\ 
\bottomrule
\end{tabular}
\caption{Bidding and Latency Metrics of All Builders (UltraSound Relay)}
\label{tab:all_latency}
\end{table}

\begin{landscape}
\begin{table}[htbp]
\centering
\footnotesize
\setlength{\tabcolsep}{2pt}
\begin{tabular}{lcccccccr}
\toprule
& Excluded & Excluded  & Excluded & Other Fee Payer  & Promise Delivered  & Over-Promised  & Under-Promised  & Relay \\
Builder &  Blocks  & Blocks  &  Value & Blocks  & Blocks  & Bid Value & Bid Value & Payment \\ 
\ & [\#] & [\%] & [ETH] & [\%] & [\%] & [ETH] & [ETH] & [ETH] \\ 
\midrule
beaverbuild & \SI{1231.0}{} & \SI{0.3}{} & \SI{45.67}{} & \SI{5.08}{} & \SI{99.66}{} & \SI{92.8}{} & \SI{-14.0}{} & \SI{4.81}{} \\ 
rsync & \SI{588.0}{} & \SI{0.21}{} & \SI{17.51}{} & \SI{1.88}{} & \SI{99.75}{} & \SI{55.93}{} & \SI{-5.88}{} & \SI{0.39}{} \\ 
Titan & \SI{92879.0}{} & \SI{34.3}{} & \SI{8412.7}{} & \SI{7.35}{} & \SI{99.66}{} & \SI{55.27}{} & \SI{-4.72}{} & \SI{1.29}{} \\ 
flashbots & \SI{108.0}{} & \SI{0.14}{} & \SI{3.31}{} & \SI{0.0}{} & \SI{99.03}{} & \SI{977.38}{} & \SI{-4.03}{} & \SI{0.0}{} \\ 
builder0x69 & \SI{79.0}{} & \SI{0.22}{} & \SI{2.26}{} & \SI{0.0}{} & \SI{99.76}{} & \SI{3.7}{} & \SI{-0.38}{} & \SI{0.0}{} \\ 
jetbldr & \SI{184.0}{} & \SI{0.56}{} & \SI{21.14}{} & \SI{1.57}{} & \SI{99.66}{} & \SI{0.94}{} & \SI{-1.05}{} & \SI{0.0}{} \\ 
f1b & \SI{24.0}{} & \SI{0.14}{} & \SI{0.57}{} & \SI{0.0}{} & \SI{99.83}{} & \SI{3.3}{} & \SI{-0.13}{} & \SI{0.0}{} \\ 
Gambit Labs & \SI{1918.0}{} & \SI{11.15}{} & \SI{249.88}{} & \SI{0.0}{} & \SI{97.06}{} & \SI{416.15}{} & \SI{-0.24}{} & \SI{0.0}{} \\ 
penguin & \SI{991.0}{} & \SI{6.35}{} & \SI{47.26}{} & \SI{0.01}{} & \SI{96.38}{} & \SI{3470.56}{} & \SI{-0.22}{} & \SI{0.0}{} \\ 
tbuilder & \SI{15.0}{} & \SI{0.14}{} & \SI{0.28}{} & \SI{0.0}{} & \SI{99.84}{} & \SI{0.11}{} & \SI{-0.06}{} & \SI{0.0}{} \\ 
lokibuilder & \SI{13.0}{} & \SI{0.23}{} & \SI{0.26}{} & \SI{0.0}{} & \SI{99.72}{} & \SI{2.08}{} & \SI{-0.72}{} & \SI{0.0}{} \\ 
BuildAI & \SI{5.0}{} & \SI{0.14}{} & \SI{0.1}{} & \SI{0.0}{} & \SI{99.75}{} & \SI{0.96}{} & \SI{-0.02}{} & \SI{0.0}{} \\ 
eden & \SI{5.0}{} & \SI{0.22}{} & \SI{0.22}{} & \SI{0.0}{} & \SI{99.78}{} & \SI{0.13}{} & \SI{-0.02}{} & \SI{0.0}{} \\ 
boba-builder & \SI{2.0}{} & \SI{0.09}{} & \SI{0.03}{} & \SI{0.0}{} & \SI{99.86}{} & \SI{0.09}{} & \SI{-0.06}{} & \SI{0.0}{} \\ 
eth-builder & \SI{10.0}{} & \SI{0.52}{} & \SI{0.26}{} & \SI{0.0}{} & \SI{99.22}{} & \SI{0.65}{} & \SI{-0.02}{} & \SI{0.0}{} \\ 
Beelder & \SI{3.0}{} & \SI{0.17}{} & \SI{0.08}{} & \SI{0.0}{} & \SI{99.49}{} & \SI{0.06}{} & \SI{-0.15}{} & \SI{0.0}{} \\ 
Anon:0x83bee & \SI{2.0}{} & \SI{0.12}{} & \SI{0.05}{} & \SI{0.0}{} & \SI{99.69}{} & \SI{0.72}{} & \SI{-0.04}{} & \SI{0.0}{} \\ 
bloXroute & \SI{1598.0}{} & \SI{58.49}{} & \SI{245.56}{} & \SI{0.0}{} & \SI{99.38}{} & \SI{0.33}{} & \SI{-0.05}{} & \SI{0.0}{} \\ 
Anon:0xb3a6d & \SI{0.0}{} & \SI{0.0}{} & \SI{0.0}{} & \SI{0.0}{} & \SI{100.0}{} & \SI{0.0}{} & \SI{0.0}{} & \SI{0.0}{} \\ 
BTCS & \SI{7.0}{} & \SI{0.73}{} & \SI{0.22}{} & \SI{0.0}{} & \SI{97.61}{} & \SI{0.11}{} & \SI{-0.26}{} & \SI{0.0}{} \\ 
I can haz block? & \SI{0.0}{} & \SI{0.0}{} & \SI{0.0}{} & \SI{0.0}{} & \SI{100.0}{} & \SI{0.0}{} & \SI{0.0}{} & \SI{0.0}{} \\ 
payload & \SI{1.0}{} & \SI{0.23}{} & \SI{0.02}{} & \SI{0.0}{} & \SI{99.53}{} & \SI{0.01}{} & \SI{0.0}{} & \SI{0.0}{} \\ 
bobTheBuilder & \SI{0.0}{} & \SI{0.0}{} & \SI{0.0}{} & \SI{0.0}{} & \SI{100.0}{} & \SI{0.0}{} & \SI{0.0}{} & \SI{0.0}{} \\ 
lightspeedbuilder & \SI{28.0}{} & \SI{7.31}{} & \SI{7.18}{} & \SI{0.0}{} & \SI{99.74}{} & \SI{0.18}{} & \SI{0.0}{} & \SI{0.0}{} \\ 
EigenPhi & \SI{0.0}{} & \SI{0.0}{} & \SI{0.0}{} & \SI{0.0}{} & \SI{100.0}{} & \SI{0.0}{} & \SI{0.0}{} & \SI{0.0}{} \\ 
nfactorial & \SI{1.0}{} & \SI{0.3}{} & \SI{0.02}{} & \SI{0.0}{} & \SI{99.7}{} & \SI{0.05}{} & \SI{0.0}{} & \SI{0.0}{} \\ 
smithbuilder & \SI{0.0}{} & \SI{0.0}{} & \SI{0.0}{} & \SI{0.0}{} & \SI{100.0}{} & \SI{0.0}{} & \SI{0.0}{} & \SI{0.0}{} \\ 
panda & \SI{0.0}{} & \SI{0.0}{} & \SI{0.0}{} & \SI{0.0}{} & \SI{100.0}{} & \SI{0.0}{} & \SI{0.0}{} & \SI{0.0}{} \\ 
Ty For The Block & \SI{1.0}{} & \SI{0.36}{} & \SI{0.02}{} & \SI{0.0}{} & \SI{99.64}{} & \SI{0.08}{} & \SI{0.0}{} & \SI{0.0}{} \\ 
antbuilder & \SI{2.0}{} & \SI{0.96}{} & \SI{0.05}{} & \SI{0.0}{} & \SI{99.04}{} & \SI{0.59}{} & \SI{0.0}{} & \SI{0.0}{} \\ 
merkle Block & \SI{0.0}{} & \SI{0.0}{} & \SI{0.0}{} & \SI{0.0}{} & \SI{100.0}{} & \SI{0.0}{} & \SI{0.0}{} & \SI{0.0}{} \\ 
ibuilder & \SI{0.0}{} & \SI{0.0}{} & \SI{0.0}{} & \SI{0.0}{} & \SI{100.0}{} & \SI{0.0}{} & \SI{0.0}{} & \SI{0.0}{} \\ 
s0e2t & \SI{1.0}{} & \SI{0.81}{} & \SI{0.03}{} & \SI{0.0}{} & \SI{99.19}{} & \SI{0.14}{} & \SI{0.0}{} & \SI{0.0}{} \\ 
ashpool & \SI{0.0}{} & \SI{0.0}{} & \SI{0.0}{} & \SI{0.0}{} & \SI{100.0}{} & \SI{0.0}{} & \SI{0.0}{} & \SI{0.0}{} \\ 
smithbot & \SI{2.0}{} & \SI{4.88}{} & \SI{0.06}{} & \SI{0.0}{} & \SI{92.68}{} & \SI{0.02}{} & \SI{-0.07}{} & \SI{0.0}{} \\ 
If you build it, they will come. & \SI{0.0}{} & \SI{0.0}{} & \SI{0.0}{} & \SI{0.0}{} & \SI{100.0}{} & \SI{0.0}{} & \SI{0.0}{} & \SI{0.0}{} \\ 
wenmerge & \SI{0.0}{} & \SI{0.0}{} & \SI{0.0}{} & \SI{0.0}{} & \SI{100.0}{} & \SI{0.0}{} & \SI{0.0}{} & \SI{0.0}{} \\ 
Manifold & \SI{320.0}{} & \SI{99.38}{} & \SI{12.65}{} & \SI{0.0}{} & \SI{91.61}{} & \SI{0.22}{} & \SI{-2.26}{} & \SI{0.0}{} \\ 
Builder Boi & \SI{0.0}{} & \SI{0.0}{} & \SI{0.0}{} & \SI{0.0}{} & \SI{100.0}{} & \SI{0.0}{} & \SI{0.0}{} & \SI{0.0}{} \\ 
\bottomrule
\end{tabular}
\caption{Validator Payment Metrics of All Builders}\label{tab:validator_payment_patterns}
\end{table}
\end{landscape}

\end{document}